\documentclass[iop,twocolappendix,numberedappendix,apj]{emulateapj}
\usepackage{epsfig, natbib, graphicx, color, amsmath, cancel,icomma,bm}

\newcommand{\Mpc}{\mbox{Mpc}}

\newcommand{\avg}[1]{\left\langle #1 \right\rangle}

\newcommand{\zmax}{z_{\mathrm{max}}}

\newcommand{\be}{\begin{equation}}
\newcommand{\ee}{\end{equation}}
\newcommand{\bea}{\begin{eqnarray}}
\newcommand{\eea}{\end{eqnarray}}

\newcommand{\kpc}{\mbox{kpc}}

\newcommand{\keV}{\mbox{keV}}
\newcommand{\Ysz}{Y_{\mathrm{SZ}}}

\newcommand{\chandra}{{\it Chandra}}

\newcommand{\Mgas}{M_{\mathrm{gas}}}

\newcommand{\zlambda}{z_\lambda}
\newcommand{\zref}{z_{\mathrm{ref}}}
\newcommand{\zmcxc}{z_{\mathrm{MCXC}}}
\newcommand{\Rmax}{R_{\mathrm{max}}}
\newcommand{\zphoto}{z_{\mathrm{photo}}}
\newcommand{\Tmin}{T_{\mathrm{min}}}
\newcommand{\lambdamin}{\lambda_{\mathrm{min}}}

\newcommand{\zspec}{z_{\mathrm{spec}}}
\newcommand{\lpivot}{\lambda_{\mathrm{pivot}}}

\newcommand{\redmapper}{{\rm redMaPPer}}
\newcommand{\photoz}{photo-$z$}
\newcommand{\photozs}{photo-$z$s}

\shortauthors{Rozo \& Rykoff}
\shorttitle{\redmapper{} II: X-ray and SZ Performance Benchmarks}

\begin{document}

\title{redMaPPer II: X-ray and SZ Performance Benchmarks for the SDSS Catalog}
\author{E.~Rozo\altaffilmark{1}, E.~S.~Rykoff\altaffilmark{1}
  }

\altaffiltext{1}{SLAC National Accelerator Laboratory, Menlo Park, CA 94025.}

\begin{abstract}
We evaluate the performance of the SDSS DR8 \redmapper{} photometric cluster
catalog by comparing it to overlapping X-ray and SZ selected
catalogs from the literature. We confirm the \redmapper{} photometric redshifts are nearly unbiased
($\avg{\Delta z} \leq 0.005$), have low scatter ($\sigma_z \approx 0.006-0.02$, depending
on redshift), and have a low catastrophic failure rate ($\approx 1\%$).
Both the $T_X$--$\lambda$ and $\Mgas$--$\lambda$ scaling relations are consistent
with a mass scatter of $\sigma_{\ln M|\lambda}\approx 25\%$, albeit with a $\approx 1\%$
outlier rate due to projection effects.   This failure rate is somewhat lower than that expected
for the full cluster sample, but is consistent with the additional selection effects 
introduced by our reliance on X-ray and SZ selected reference cluster samples.
Where the \redmapper{} DR8 catalog is volume limited ($z\leq 0.35$),
the catalog is 100\% complete above $T_X \gtrsim 3.5\ \keV$,
and $L_X \gtrsim 2\times 10^{44} \mathrm{erg}\,\mathrm{s}^{-1}$, decreasing to 90\% 
completeness at $L_X \approx 10^{43}\ \mathrm{erg}\,\mathrm{s}^{-1}$.
All rich ($\lambda \gtrsim 100$), low redshift ($z\lesssim 0.25$) \redmapper{} 
clusters are X-ray detected in the ROSAT All Sky Survey (RASS), and $86\%$ of the clusters
are correctly centered.  Compared to other SDSS photometric cluster catalogs,
\redmapper{} has the highest completeness and purity, and the best photometric redshift performance, 
though some algorithms do achieve comparable performance to \redmapper\
in subsets of the above categories and/or in limited redshift ranges.   The \redmapper\ richness is clearly the one
that best correlates with X-ray temperature and gas mass.  Most algorithms (including \redmapper) have very similar 
centering performance, with only one exception which performs worse.
\end{abstract}

\keywords{galaxies: clusters}

\section{Introduction}

Galaxy clusters are well known cosmological 
probes \citep[e.g.][]{henryetal09,vikhlininetal09b,mantzetal10a,rozoetal10a,clercetal12,bensonetal13,hasselfieldetal13}.
Indeed, galaxy clusters are a key component of the current efforts to probe our accelerating Universe
with photometric surveys such as the Dark Energy Survey \citep[DES, ][]{des05}, Pan-STARRS \citep{panstarrs02}, the 
Hyper-Suprime Camera\footnote{http://www.naoj.org/Projects/HSC/HSCProject.html}, and the
Large Synoptic Survey Telescope \citep[LSST, ][]{lsst12}.   For these efforts to succeed, one requires a robust, well calibrated, fully
optimized photometric cluster finding algorithm.

This work is a companion paper to Rykoff et al. (2013, henceforth Paper I), which introduces \redmapper, a new optical cluster finding
algorithm specifically designed to take full advantage of these upcoming photometric surveys.  
The primary goal of this work is to evaluate the performance of \redmapper\ against a variety of X-ray
and SZ selected catalogs.  The motivation for this type of comparison is manifold. 

First, while it is true that galaxy clusters were first discovered in the optical as far back as the 1800's \citep{biviano00},
and that the first large cluster catalogs relied on optical observations \citep[e.g.][]{abell58,zwickyetal68,aco89}, ever since the 1990's photometric
cluster selection has been viewed as relatively unreliable.  In particular, optical selection is broadly considered
to be prone to severe projection 
effects \citep{lucey83,frenketal90,vanhaarlemetal97}.  While these early results were concerned with photometric
cluster finding with single band imaging, the modern literature has clearly demonstrated that projection effects remain a
critical concern for optically selected samples, even with the advent of multi-band data \citep[e.g.][]{cohnetal07,whiteetal10,nohcohn11,anguloetal12}.  
While it has been argued that projection effects can be minimized through a careful treatment of photometric data \citep[][Paper I]{rozoetal11a},
it remains to be explicitly demonstrated that a low incidence of projection effects can be achieved.  
Because X-ray cluster selection is extremely robust to projection effects, a direct comparison of the 
\redmapper\ catalog to X-ray selected systems and sources in the ROSAT All Sky Survey \citep[][]{vogesetal99}
allows us to directly test the incidence of severe projection 
effects in \redmapper.

Second, it has long been recognized that photometric mass proxies such as total galaxy counts
and/or total cluster luminosity may not be very effective.  That is, while these quantities are clearly correlated
with cluster mass, they appear to exhibit large scatter \citep{gladdersetal07,rozoetal09a,songetal12}.
However, recent work has argued that optimized redshift estimators can deliver low 
scatter \citep{rozoetal09b,rykoffetal12}, although the arguments have been based on noisy data with an intrinsically
large scatter \citep[though see also][]{andreon12}.  
Here, we wish to explicitly test whether a scatter in mass at fixed richness as low
as has been claimed in the
literature can be realized by comparing the \redmapper{} richness estimator against low-scatter X-ray mass proxies such as X-ray
temperature and gas mass.

Third, it is sometimes suggested that optically selected clusters represent a fundamentally distinct population of galaxy
clusters from X-ray and/or SZ selected galaxy clusters --- the so-called X-ray under-luminous systems --- as opposed to 
X-ray and optically selected cluster samples
having regular scaling relations \citep[e.g.,][]{donahueetal01,rasmussenetal06,lopesetal09,daietal10,andreonmoretti11,
baloghetal11,planck11_optical,wangetal11,rozoetal12d}.  
Of course, in detail, no optical cluster catalog can be
precisely the same as an X-ray or SZ selected catalog.  Rather, the relevant question is whether the relative
selection functions behave as expected.  For instance, optically selected clusters cannot be detected in X-rays
if they are fainter than the flux limit of the X-ray survey, so one naturally expects the fraction of optically selected
clusters detected in X-rays to decline in a simple, predictable manner with both richness and redshift.  Is this, in fact,
the case?  Similarly, are there any massive X-ray and/or SZ selected clusters which one would expect to be
detected in optical, but which are not?  Addressing these questions if of critical importance in cementing 
photometric cluster selection as a viable alternative to X-ray and SZ cluster surveys.

Of course, \redmapper\ is far from the only cluster finding algorithm available in the literature today.
At present, there exists a large variety of photometric cluster finding algorithms, with roughly half of the cluster finders
relying on photometric galaxy redshifts for cluster identification \citep[e.g.,][]{kepneretal99,vanbreukelenetal09,milkeraitisetal10,
durretetal11,spdpg11,soaressantosetal11,wenetal12} and half (including \redmapper) relying on the red-sequence 
technique
\citep{annisetal99,gladdersyee00,koesteretal07a,gladdersetal07,galetal09,thanjavuretal09,haoetal10}.
Given the variety of different algorithms and techniques employed in optical cluster finding, maximizing
the utility of ongoing and near future photometric surveys necessitates a detailed comparison of these various 
algorithms.   In this way, we may adequately identify which techniques are better or worse suited to the specific
question at hand, in particular cluster cosmology.   The secondary goal of this work then is to compare and contrast the performance of \redmapper\ against that of
several other cluster finders.  

The basic analysis framework that we use for the comparison of \redmapper\ to other cluster finding algorithms
is fundamentally different than the type of cross-comparison work that has been performed in the past.  Specifically, 
we do not directly compare \redmapper\ against the other catalogs.  The reason is simple: 
consider, for instance, comparing two photometric cluster catalogs $A$ and $B$.  If the richness comparison
of catalogs $A$ and $B$ is noisy, how can one tell whether $A$ is at fault, or whether $B$ is at fault?  If a specific
cluster is found in catalog $A$, but not catalog $B$, does that constitute a failure of catalog $A$, a failure of
catalog $B$, or neither, or both?  Because one does not know {\it a priori} which catalog is ``more correct'' --- and this
needs to be properly defined --- it is unclear
what kind of inferences could be made from such a comparison \citep[see e.g.,][for further discussion]{bahcalletal03b}.
What one needs to perform a proper comparison then is some sort of ``truth table'', i.e., a definition of ``more correct''.

In this work, we rely on X-ray and SZ selected catalogs as a truth-table of sorts.  So, for instance, given two optical catalogs
$A$ and $B$, we can unambiguously determine whether the fraction of galaxy clusters detected in X-rays is higher for 
catalog $A$ than for catalog $B$ or vice versa.  Provided galaxy clusters form a single population with well defined
scaling relations,  then a larger fraction
of X-ray selected clusters can be unambiguously viewed as a good thing.  Similarly,
a lower scatter in X-ray temperature and/or gas mass at fixed richness is also clearly a good thing, as both of these
quantities are recognized as very accurate mass proxies.  

In addition, it is our hope that as various features of photometric cluster
finding algorithms are revealed to be more or less effective, that the field as a whole will converge into a more unified framework.
As noted above, there are now at least 15 different
cluster finding algorithms, all of which have different selection functions, and different richness and redshift estimators.
Consequently, it is difficult to compare the results from various research groups, even more so given that each
cluster finding algorithm is typically run on only one data set.  By identifying which techniques are more or less
effective, we hope the field can converge into a common language that will facilitate communication.   Conversely,
if we find that several techniques are all found equally effective, then there is a strong motivation to pursue multiple
cluster detection avenues in future work, so as to provide an explicit test of cluster cosmology with several
independent samples.   

Our work is laid out as follows.  In Section \ref{sec:cats}, we introduce the \redmapper\ catalog, and the
various X-ray and SZ catalogs that will be employed in our analysis of the performance of \redmapper. 
Section \ref{sec:matching} discusses how we match galaxy clusters across catalogs, a process that
is the fundamental stepping stone for all subsequent work.  Section \ref{sec:photoz} compares the
\redmapper\ redshifts to those of our reference catalogs, while Section \ref{sec:scalings} 
explore the X-ray and SZ scaling relations with the \redmapper\ optical richness.  
Section \ref{sec:comp_pur} discusses the relative X-ray and SZ completeness and purity of the
\redmapper\ catalog, and section \ref{sec:centering} evaluates the efficacy of the \redmapper\ 
centering algorithm.  Finally, in Section \ref{sec:optical_comparison} we perform a simplified
version of all of the above analyzes for all photometric cluster catalogs available in the
SDSS (including \redmapper).  While this analysis is less thorough than that 
used for the detailed characterization of \redmapper\ --- in particular we do not perform
extensive visual inspection of all cluster catalogs and their matches to the X-ray and SZ 
catalogs --- the analysis is completely homogeneous: all cluster catalogs are treated in
exactly the same way.    Section \ref{sec:summary} summarizes our results and present
our conclusions.
When necessary, distances are estimated assuming a flat $\Lambda$CDM model with $\Omega_m=0.27$, and $h=0.7\ \Mpc$.


\section{Catalogs}
\label{sec:cats}

\subsection{The SDSS DR8 redMaPPer Catalog}
\label{sec:redmapper}

The {\bf red}-sequence {\bf Ma}tched-filter {\bf P}robabilistic {\bf Per}colation (\redmapper)
algorithm~(Paper I: Rykoff et al. 2013) is a photometric cluster finding algorithm
based on the optimized richness
estimator $\lambda$~\citep{rozoetal09b,rykoffetal12}.  \redmapper{}
identifies galaxy clusters as overdensities of
red-sequence galaxies.  It relies on an iterative self-training technique to
fully characterize the evolution of the red sequence as a function
of redshift, including zero-point, tilt,
and scatter.  As the algorithm utilizes all colors ($u-g$, $g-r$, $r-i$, and
$i-z$) simultaneously, the ``scatter'' is characterized by a full covariance
matrix.  The algorithm then uses the resulting red sequence model,
combined with simple radial
and luminosity filters, to estimate the probability that any given galaxy
belongs to any given cluster.  The cluster richness $\lambda$ is 
the sum of these probabilities.  In addition, the algorithm estimates
cluster photometric redshifts by simultaneously fitting all high probability
cluster members with a single red sequence model.

The \redmapper{} catalog used in this work is that obtained from running
the algorithm on the SDSS DR8
data~\citep{dr8}.  The survey mask is the same as that used for the Baryon
Acoustic Oscillation Survey (BOSS) target selection~\citep{dsaaa13},
supplemented by some additional cuts around unmasked bright 
stars and foreground galaxies (e.g. M31), for a total area of $\approx 10,000\ \deg^2$.
All richness estimates are corrected for masked
area due to survey edges, bright stars, and bad fields, and the catalog is further trimmed
so that no cluster is masked by more than $20\%$.
Spectroscopic redshifts used for photometric
redshift training and validation are derived from a compilation of SDSS
main~\citep{straussetal02}, luminous red
galaxy~\citep[LRG,][]{eisensteinetal01}, and BOSS DR9~\citep{dr9} galaxy
samples.  For a detailed description of how the photometric and spectroscopic
catalogs are used in the training and construction of the catalog, we refer the
reader to Paper I.


\subsection{X-ray and SZ Catalogs}

The SDSS DR8 \redmapper\ catalog will be compared to the X-ray and
SZ catalogs described below.  In
all cases, we adopt a cut on data quality: if $X$ is the cluster
observable of interest (e.g. $L_X$, $T_X$, $\Mgas$), we discard from the
reference cluster catalog any systems with $\Delta X/X \geq 0.3$.  
The catalogs employed in our analysis are:
\\

\noindent {\bf XCS:} The XMM Cluster Survey \citep[XCS:][]{mehrtensetal12}
is a serendipitous search for galaxy clusters using all publicly available data in the {\it XMM-Newton} Science
Archive.  The first data release is comprised of 503 optically confirmed, serendipitously detected galaxy clusters.  Of these,
261 have spectroscopic redshifts, and 203 have photometric redshift estimates.  We restrict ourselves to the sub-sample
of 402 galaxy clusters with temperature estimates.   By necessity, the X-ray temperatures reported in the XCS catalog
are not core-excised. 
\\

\noindent {\bf MCXC:} The Meta-Catalog of X-ray detected Clusters of galaxies
\citep[MCXC:][]{piffarettietal11} is a compilation of galaxy clusters based on
publicly available X-ray data from both the
ROSAT All Sky Survey~\citep[RASS:][]{vogesetal99} and serendipitous searches
in ROSAT pointed observations.  The RASS contributing catalogs are
NORAS~\citep{bohringeretal00}, REFLEX~\citep{bohringeretal04}, BCS~\citep{ebelingetal98}, 
SGP~\citep{cvbcr02}, NEP~\citep{hmvbb06}, MACS~\citep{eeh01}, and CIZA~\citep{emt02,kemt07},
while the contributing serendipitous catalogs are 160D~\citep{mmqvh03},
400D~\citep{bvheq07}, SHARC~\citep{rnhup00}, WARPS~\citep{hpejs08}, and
EMSS~\citep{gmsws90}.  The data from each of the individual galaxy catalogs was
collected and homogenized, deleting duplicate entries, and enforcing a
consistent X-ray luminosity definition.  In the catalog, $L_X$ is defined to be the
X-ray luminosity in the 0.1--2.4~keV band within an $R_{500c}$ aperture.  Unfortunately, the catalog
does not include errors in the X-ray luminosity estimates.
\\

\noindent {\bf ACCEPT:} The ACCEPT cluster catalog is a compilation of X-ray clusters with deep \chandra\ data \citep{cavagnoloetal09}.
All data were independently reduced and homogeneously analyzed, with projected temperature radial profiles
available for all galaxy clusters.  However, a single core-excised X-ray temperature is not reported.  
We utilize the temperature and gas profiles to compute a spectroscopic-like core-excised temperature using the
weighting scheme in \citet{mazzottaetal04}.  The necessary integrals from \citet{mazzottaetal04} are discretized into a
sum over the observed radial bins, and
core-excision is done using a $150\ \kpc$ aperture when possible.
If no radial bins falls entirely outside this region, we do not perform 
core-excision.

Errors in the X-ray temperatures
are estimated via direct Monte Carlo: the density and temperature profiles are randomly sampled based on their
reported errors, and these are used to compute the average temperature as described above.  The error is defined to be the standard
deviation of the Monte Carlo samples for each cluster.  Because ACCEPT reports their interpolated temperature profiles (so as
to match their gas density profiles)  there is significant covariance between neighboring radial bins.  This leads to an under-estimate
of the true temperature uncertainties in our Monte Carlo method.  Unfortunately, the non-interpolated profiles were not available,
so we simply
assume that 1-off neighboring bins are perfectly correlated, with no covariance between non-neighboring
bins.  This increases our estimated error by a factor of $\sqrt{2}$, though we hasten to add that these errors
have a negligible impact on the recovered scatter in X-ray temperature
at fixed richness.

There is one cluster in this catalog that deserves special mention. Specifically,  visual inspection of the density and temperature profiles 
of Abell 1942 reveals an obvious failure in the automated data reduction for $R \geq 500\ \kpc$, so including the information in the reported
profiles beyond these radius severely biases the recovered X-ray temperature.  Consequently, when estimating $T_X$ for this system,
we truncate all profiles at $R=400\ \kpc$.  We have explicitly verified with the ACCEPT team that truncating the integration at $R=400\ \kpc$ for this galaxy cluster
is appropriate, and that failure to do so will introduce large systematic errors in the recovered X-ray temperature (M. Donahue 2012, private communication).
We note that if we do not truncate the profiles at this radius for Abell 1942, then this cluster becomes a gross outlier in the $T_X$--$\lambda$
relation (see section \ref{sec:scalings}).
\\

\noindent {\bf Mantz:} The Mantz cluster sample~\citep{mantzetal10b} is comprised of galaxy clusters contained within the ROSAT Brightest Cluster Sample \citep[BCS:][]{ebelingetal98},
the ROSAT-ESO Flux-Limited X-ray sample \citep[REFLEX:][]{bohringeretal04}, and the bright sub-sample of  the Massive Cluster Survey
\citep[Bright MACS:][]{ebelingetal10}.  Clusters from each of these catalogs were selected by applying a redshift-dependent flux cut, so as to select
the most luminous X-ray clusters in the various samples.  A sub-sample of these galaxy clusters was observed with \chandra\ in order to derive 
X-ray temperatures and gas masses. Our analysis is restricted to this sub-sample of galaxy clusters with gas mass measurements.
\\

\noindent {\bf Planck ESZ:} The Planck all-sky Early Sunyaev-Zeldovich Cluster sample \citep{planck11_earlysample} 
comprises 189 galaxy clusters detected with signal-to-noise
$S/N\geq 6$ in the Planck early data release.  Of these, 20 were newly detected
clusters, most with perturbed morphologies in the X-rays.
The clusters are typically massive low redshift clusters, with the effective
mass threshold increasing with increasing redshift. 
\\

\noindent {\bf ROSAT Bright and Faint Source Catalogs:} The ROSAT Bright
\citep{vogesetal99} and Faint \citep{vogesetal00} Source Catalogs are comprised
of all X-ray sources in the ROSAT All Sky Survey.   
X-ray detection is based on count rate in the 0.1--2.4~keV band, but sources must also
pass likelihood thresholds, and have a minimum number of photon counts.  The resulting catalog has over $10^5$ X-ray sources over the entire
sky.


\subsection{Richness and Photometric Redshift Estimates of Clusters in the Reference Catalogs}
\label{sec:measure_refs}

To facilitate cross-catalog matching, we explicitly measure the richness and
photometric redshift of every cluster in our reference catalogs that falls
within our DR8 galaxy mask.  In
doing so, we hold the center of the galaxy clusters fixed to the reported X-ray
centers.  We estimate the cluster richness $\lambda$ in exactly the same way as
we do for clusters detected in the \redmapper{} catalog, as described in Paper
I.  We note that while we hold the cluster center fixed, when estimating cluster
richness we do not rely on the
reported cluster redshifts in the reference catalogs, but rather estimate the
photometric redshift $\zlambda$ from the photometric data.  When measuring
richness, the reported
cluster redshift is only employed when initializing our iterative photometric redshift
estimate algorithm.


\section{Cluster Matching}
\label{sec:matching}

Throughout this work, we will be comparing the properties of the X-ray/SZ-selected cluster catalogs
to the \redmapper{} clusters, so we must first define an algorithm for matching galaxy clusters
between different cluster catalogs.
We consider two types of matching algorithms, namely cylindrical (or
proximity) matching, and membership matching.

\subsection{Cylindrical Matching}
\label{sec:cylindrical}

Cylindrical matching is simple: given a cluster $x\in X$, we wish to find the corresponding
cluster match $y\in Y$.  We rank order all clusters $x\in X$, and then find all clusters
$y$ within some physical radius $\Rmax$ (evaluated at the redshift $z_x$ of cluster x)
and within some maximum redshift offset $\Delta \zmax$.  Here, we set $\Rmax=1.5\ \Mpc$
and $\Delta \zmax=0.1$, and the redshift is always that reported in the original catalogs (as opposed
to our own photometric redshift estimate).  If more than 2 clusters $y\in Y$ fall within the cylinder centered
on $x$, we take the largest of the 2 as the correct match, and remove this cluster from
consideration when matching subsequent clusters in $X$.  Thus, all cluster matches are unique.

The above procedure defines the cluster match $y$ of a cluster $x$.  Using a similar algorithm,
we can then find the cluster match $\tilde x$ of the cluster $y$, which may or may not be
the same as the original cluster $x$.  Matches are considered spurious if $\tilde x \neq x$,
and are dropped from the matched cluster list.

\subsection{Membership Matching}
\label{sec:membership}

Membership matching follows the same general algorithm to enforce unique, two-way matches
between the $X$ and $Y$ cluster catalogs --- i.e. we first match Y to X and then X to Y, keeping only
clusters which are two-way matches.  The main difference is the criterion used to match a cluster
$y$ to a cluster $x$.  Specifically, let $p^x_i$ be the probability that galaxy $i$ belongs
to cluster $x\in X$, and $p^y_i$ be the probability that galaxy $i$ belongs to cluster $y\in Y$.
We define the matching strength $s(x,y)$ between two clusters $x$ and $y$ via
\be
s(x,y) = \sum_i p^x_i p^y_i.
\ee
The match $y$ to cluster $x$ is that which maximizes the function $s(x,y)$ at fixed $x$.
Likewise, the cluster match $\tilde x$ to cluster $y$ is that which maximizes $s(x,y)$
at fixed $y$.   As before, if $\tilde x \neq x$ then the match is considered spurious and dropped from the
list of matching clusters.

\subsection{Testing The Matching Algorithms}

For the vast majority of our clusters, the results from the cylindrical and
membership matching agree.  For each of our reference catalogs, we visually
inspect all clusters for which the cylindrical and membership matches disagree.
When possible, any such ambiguous cluster matchings are resolved based on
our visual inspection.  These disagreements are relatively rare, occurring in
$\lesssim 5\%$ of all clusters.  Notes for each of these systems are collected
in Appendix \ref{app:vismatching}.

Additionally, recall that we have estimated the cluster richness for every cluster in
our reference cluster catalogs.  One should expect that all such clusters that
satisfy the \redmapper\ richness cuts should be matched to an existing
cluster in the \redmapper\ catalog.   We find that there are a total of 9 galaxy
clusters across all catalogs which were unmatched.  None of these
are unmatched due to a failure of the matching algorithms.

Specifically, one cluster is unmatched because it falls within an unmasked
region in the SDSS with bad photometry.
Three of the remaining eight are unmatched because they formally
fall outside the \redmapper{} angular selection threshold --- i.e. their
location on the sky is such that more than $20\%$ of the cluster is masked
out.  The remaining 
five galaxy clusters have richnesses that are very close to the richness
threshold applied, and therefore
scatter in and out of the sample due to the small differences in redshift
and/or cluster centers between the \redmapper{} and reference catalogs.  

It is worth noting that 
these results also allow us to estimate the optical completeness, i.e. the
fraction of galaxy clusters that pass our selection threshold but are not in
the catalog due to catastrophic failures in the optical data and/or algorithm.
Since there are 372 clusters MCXC satisfying our cuts, only one of which is
rightfully unmatched due to bad photometry, 
we conclude that our optical incompleteness is $\lesssim
0.3\%$.  Notes on each of the nine clusters noted above are collected in Appendix
\ref{app:unmatched}.


\section{Redshift Comparisons}
\label{sec:photoz}

We begin by comparing the \redmapper{} photometric redshift (``\photoz'') estimator ($\zlambda$)
to the redshifts quoted in the reference catalogs.
This is a necessary first step; all cluster observables derived
using an incorrect redshift will be systematically biased, and will 
therefore compromise the study of cluster scaling relations, completeness, purity, etc.
Consequently, when comparing \redmapper{} to other catalogs,
our first task must be to identify catastrophic redshift outliers.

For each matched cluster in our reference catalogs, we calculate the redshift offset $\Delta z = \zlambda-\zref$ of
each cluster.
For the photometric redshift $\zlambda$, we consider both the photometric redshift estimate obtained while
holding the cluster center of the reference cluster fixed (see section \ref{sec:measure_refs}),
and the cluster redshift of the \redmapper{} cluster match.

The clusters are binned in redshift (based on our photometric redshift), and we require the bins be sufficiently
wide to contain at least 25 clusters per bin. 
We focus on the redshift span
$z\in[0.1,0.5]$ over which the \redmapper{} richness estimates are most robust.
At $z\geq 0.5$, richness errors become very large due to the extrapolation in luminosity between
the SDSS depth and the luminosity cut of $0.2L_*$ used to define $\lambda$.

\begin{figure*}
  \begin{center}
    \epsscale{0.6}
    \plotone{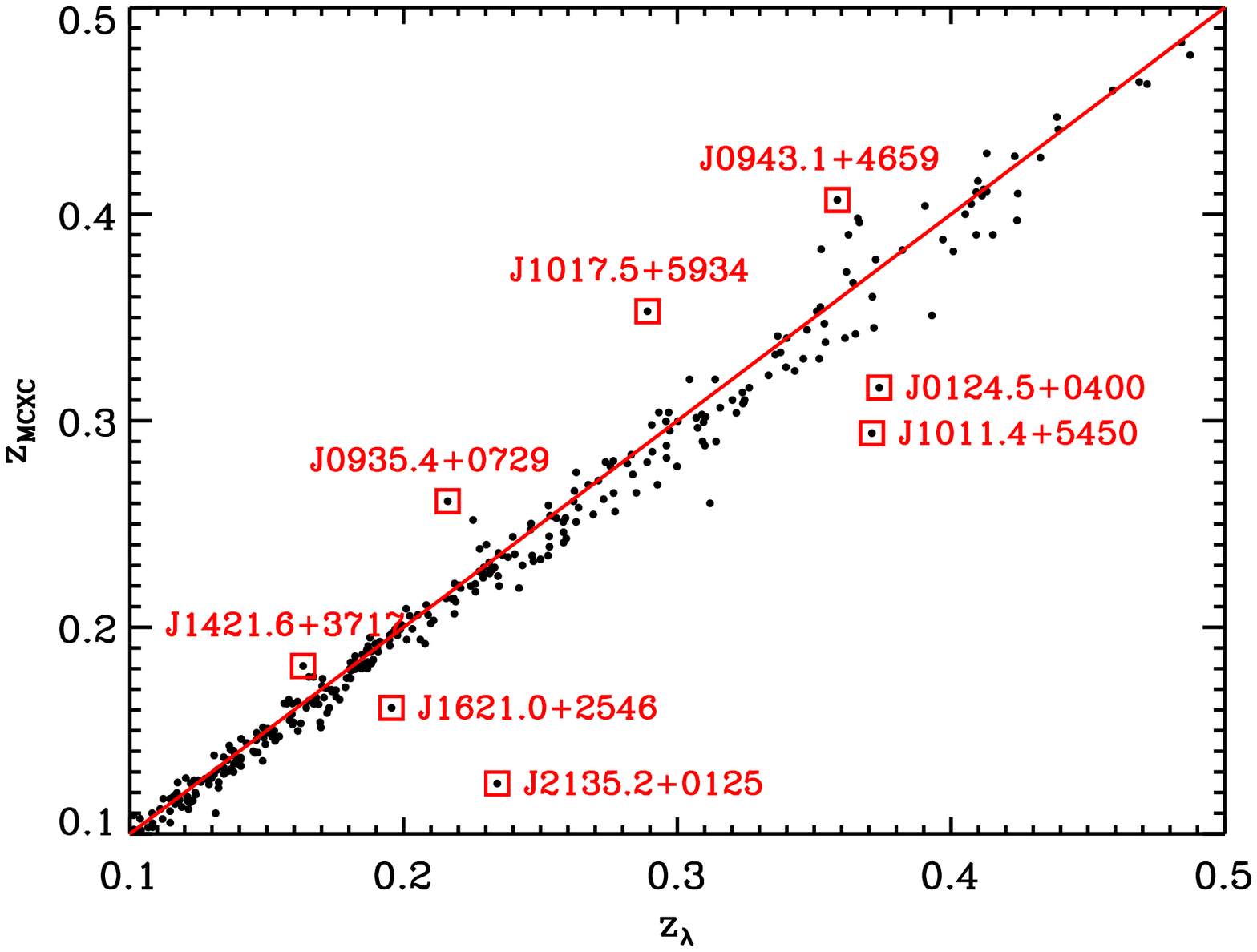} \hspace{-0.3in} \plotone{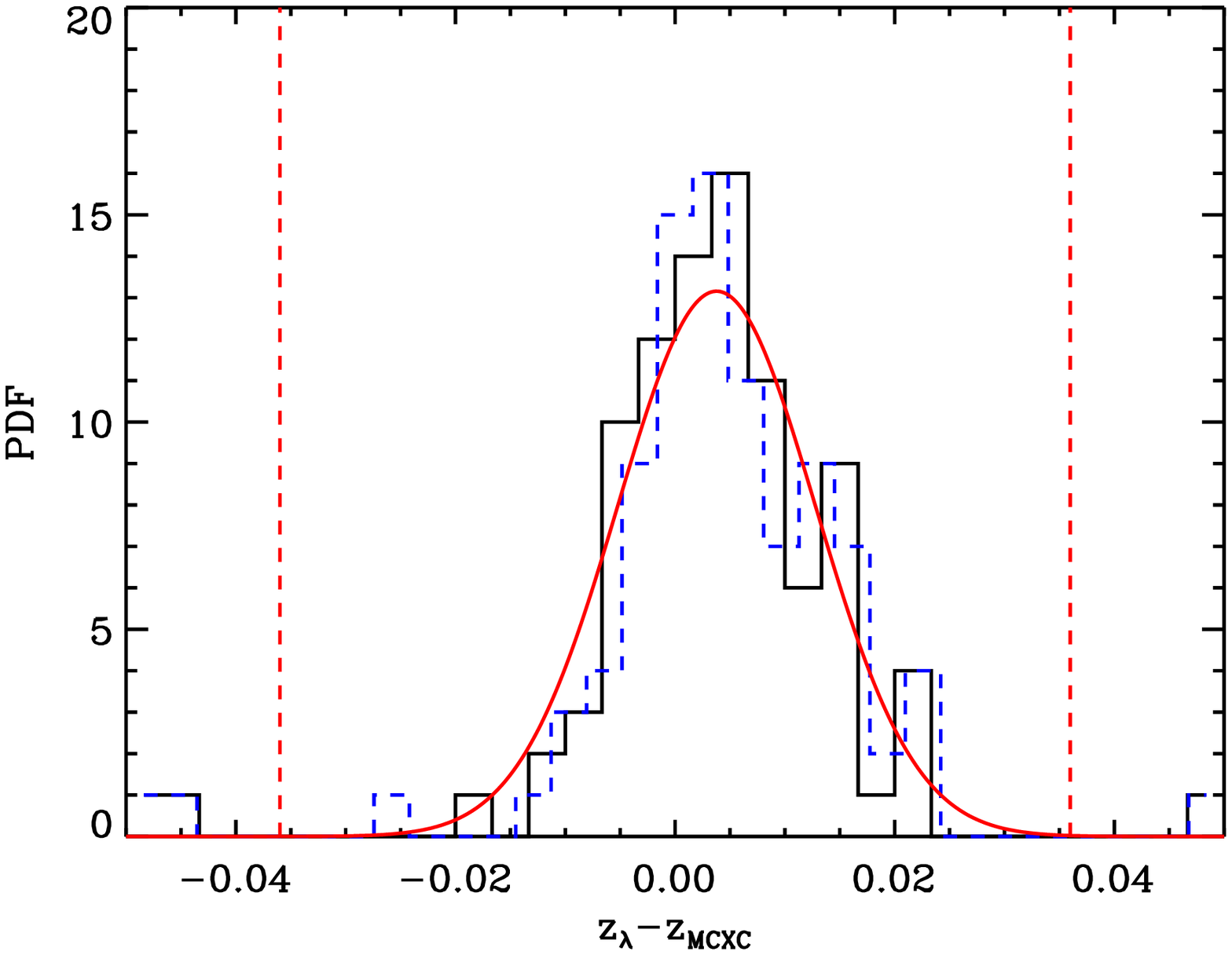}
\caption{{\it Left panel: } Comparison between the MCXC redshifts
and the redshifts of the matching \redmapper{} clusters.
Labeled clusters are flagged as outliers, and are dominated by erroneous redshifts
in the MCXC catalog (see Appendix~\ref{app:zoutliers}).
{\it Right panel: } Distribution of $\Delta z = \zlambda - \zmcxc$
in the redshift bin $\zlambda \in [0.2,0.3]$.  $\zlambda$ is either the photometric
redshift at the reported center (black histogram) or the redshift of the
matching \redmapper{} cluster (blue dashed histogram).  The apparent left--right
offset of the two histograms arises because we have utilized two slightly
different binning schemes to help differentiate between the two
histograms.
The red curve is the best fit Gaussian to the black histogram, and the
dashed red-lines mark the $4\sigma$ cut employed to identify outliers. 
}
\label{fig:photoz_dist}
\end{center}
\end{figure*}

As shown in Figure \ref{fig:photoz_dist},
the resulting distribution of redshift offsets typically exhibits
a Gaussian core with possibly a few outliers.  
The parameters for the Gaussian core are estimated
using the median redshift offset for the mean, and $1.4826\times MAD$ for the standard
deviation, where $MAD$ is the median absolute deviation of the sample about the median.\footnote{The
factor 1.4826 relates the $MAD$ to the standard deviation for a Gaussian distribution.}
All $4\sigma$ outliers are identified
and discussed.  An example of the distribution of redshift offsets for the MCXC catalog, along with the recovered Gaussian
fit and $4\sigma$ cuts is shown in Figure \ref{fig:photoz_dist}.


Our results are summarized in Table \ref{tab:photoz}.  The \photoz{} performance is consistent
with that quoted in Paper I, that is, our redshifts
are unbiased at the $\lesssim 0.005$ level, with scatter that is $\approx 0.008$ for $z\lesssim 0.3$,
but increasing to $\approx 0.02$ by $z\approx 0.5$.   
The \redmapper{} photometric failure rate is small, with two redshift failures for the full MCXC
cluster catalog, and two clusters with bad photometry.  The corresponding total failure
rate (including bad photometry) is $\approx 1.2\% \pm 0.6\%$.  The \photoz{} performance 
is largely insensitive to the choice of cluster center, though differences can arise if cluster
centers are separated by large ($R\sim 1\ \Mpc$) offsets.
Comments for each of the galaxy clusters identified as outliers can be found in Appendix \ref{app:zoutliers}.


\begin{deluxetable*}{llllccc}
  \tablewidth{0pt}
  \tablecaption{Comparison of \redmapper{} Photo-$z$s to Reference Catalogs
    \label{tab:photoz}
  }
  \tablehead{
    \colhead{Catalog} &
    \colhead{Redshift Bin} &
    \colhead{$\avg{\zlambda - \zref}$} &
    \colhead{$\sigma_z$} & 
    \colhead{Tot. Cl.} & 
    \colhead{Bad $\zref$} &
    \colhead{Bad $\zlambda$}
  }
\startdata
XCS: fixed center & [0.1,0.3] & 0.0038 & 0.0089 & 41 & 1 & ---  \\
	& [0.3,0.5] & 0.0064 & 0.0192 & 44 & --- & ---  \\
	
XCS: matched 	& [0.1,0.3] & 0.0043 & 0.0120 & 41 & 1 & ---\\
	& [0.3,0.5] & 0.0067 & 0.0206 & 44 & --- & ---  \vspace{0.04in}	\\

\hline
\vspace{-0.04in}	\\

MCXC: fixed center & [0.1,0.2] & 0.0021 & 0.0057 & 163 & 1 & --- \\
	& [0.2,0.3] & 0.0037 & 0.0089 & 91 & 3 & --- \\
	& [0.3,0.5] & 0.0053 & 0.0170 & 75 & 2 & ---  \\

MCXC: matched & [0.1,0.2] & 0.0023 & 0.0050 & 163 & 2 & --- \\
	    & [0.2,0.3] & 0.0041 & 0.0083 & 91 & 3 & ---  \\
	    & [0.3,0.5] & 0.0053 & 0.0170 & 75 & 2 & 2  \vspace{0.04in} \\

\hline
\vspace{-0.04in}	\\

ACCEPT: fixed center & [0.1,0.5] & 0.0055 & 0.0094 & 56 & 2 & --- \\
ACCEPT: matched  & [0.1,0.5] & 0.0050 & 0.0080 & 56 & 2 & ---\vspace{0.04in} \\

\hline
\vspace{-0.04in}	\\

Mantz: fixed center & [0.1,0.3] & 0.0043 & 0.0067 & 31 & --- & 1  \\
Mantz: matched  & [0.1,0.3] & 0.0031 & 0.0057 & 31 & --- & 1\vspace{0.04in} \\

\hline
\vspace{-0.04in}	\\

Planck ESZ: fixed center & [0.1,0.5] & 0.0030 & 0.0092 & 38 & 1 & ---  \\
Planck ESZ: matched & [0.1,0.5] & 0.0025 & 0.0081 & 38 & 1 & ---
\enddata
\tablecomments{ ``Fixed center'' means the photometric redshift was estimated while holding the cluster
center fixed to that reported in the reference catalog.
``Matched'' means we compare the redshift in the reference catalog to that of the matching \redmapper{} cluster.
$\sigma_z$ is the width of the
Gaussian core of the $P(\Delta z)$ distribution. ``Tot. Cl.'' is the total number of clusters in the redshift
bin.   ``Bad $\zref$'' is the number
of $4\sigma$ outliers due to erroneous redshifts in the reference catalog.  
``Bad $\zlambda$'' is the number of erroneous \redmapper{} redshifts.
The statistics in this table do not include two \redmapper{} clusters that were
compromised due to bad SDSS photometry.}
\end{deluxetable*}



\section{X-ray and SZ Scaling Relations}
\label{sec:scalings}

\subsection{Methods}

We now consider the efficacy of the \redmapper{} richness $\lambda$ as a mass tracer by
looking at cluster scaling relations.  To decouple this analysis from the performance of 
our optical centering algorithm, we rely
on the cluster richness measurements 
at the reported cluster center in the reference catalogs.  
Cluster centering will be addressed in Section \ref{sec:centering}.
Because both richness and X-ray observables depend on redshift, we remove
all redshift outliers from this analysis.  These are dominated by erroneous redshifts
in the reference catalogs, but include four \redmapper{} clusters.  Note that the richness
measurements are evaluated at our estimated photometric redshift, and not the reported
cluster redshifts of the reference catalogs, so noise associated with photometric redshift
uncertainties are included in our analysis.
Finally, we impose two data cuts: the first is that the cluster richness must pass the \redmapper{}
selection criteria, and the second is that the error in the external observable ($L_X$, $T_X$, etc)
must be less than $30\%$.  
The latter cut impacts primarily
XCS galaxy clusters with relatively few photon counts.

The mean and scatter of the X-ray scaling relations
are estimated using a standard Bayesian fitter \citep[e.g.,][]{rozoetal12a,kelly07}
with a uniform prior on the amplitude, slope, and variance of the relation.
These fits are performed in redshift bins, which are chosen to match those in
Section \ref{sec:photoz}.
Errors are estimated via bootstrap resampling.

To check for possible systematic failures of \redmapper, and/or identify
unique galaxy clusters, we have also implemented an automated outlier rejection
criteria.  We reject from our fit all $3\sigma$ outliers in the scaling relation.
There are only two such outliers, which are discussed below.
For now, we simply note that we believe that exclusion of these clusters
from our fits is well justified.

Of particular interest to us are estimates of the scatter in mass at fixed richness,
which can be inferred from the scatter in X-ray/SZ observables at fixed richness
as laid out in the Appendix of \citet{rozoetal12c}.
For simplicity, we assume zero intrinsic covariance between
the various X-ray and SZ observables and the cluster richness.
Setting the correlation coefficient $r$ between $\lambda$ and the additional observable $X$
to $r=0$ in Eq. A13 of \citet{rozoetal12c}, we find that 
the scatter in mass at fixed richness is given by
\be
\sigma_{M|\lambda}^2 = \frac{ \sigma_{X|\lambda}^2 }{\alpha_{X|M}^2 } - \sigma_{M|X}^2 
\ee
where $\alpha_{X|M}$ is the slope of the observable--mass relation for the reference catalog.
If $r\neq 0$, the scatter can be larger or smaller depending on the sign of $r$. 
For estimating the mass scatter, we adopt a fiducial
scatter in mass at fixed $M_{gas}/T_X/L_X$ of $10\%/15\%/25\%$ and
slopes of $\alpha=1.1$, $1.5$, and $1.6$. These numbers are characteristic of the vast
literature on cluster scaling relations \citep[e.g. ][and many others]{vikhlininetal09,mantzetal10b,prattetal09,
mahdavietal12,rozoetal12d}.

We emphasize that {\it the scaling relations presented in this work have not been corrected for
selection effects.}
This is because of two important considerations: first, our immediate goal is to quickly
evaluate the performance of \redmapper, leaving precise estimates of cluster scaling
relations for future works (e.g. Greer et al, in preparation).  Most importantly, though,
the selection functions of our reference catalogs are often either difficult to quantify
without the tools employed in the construction of these catalogs (e.g. XCS, Planck ESZ), or downright
impossible to quantify because the catalogs are aggregates from multiple sources 
(e.g. MCXC and ACCEPT).  For the remaining Mantz catalog, our reliance on $\Mgas$
rather than $L_X$ as the mass tracer implies that a proper treatment needs to account for the obvious covariance
between these two X-ray observables, an analysis which is beyond the scope of this work.

In this context, it is important to note that systematic offsets in our recovered
scaling relations scale with the variance of the relation.  Thus, as long as this scatter
is small, our recovered relations should be relatively robust to selection effects.
As we will see below, both the $T_X$--$\lambda$ and $\Mgas$--$\lambda$ scaling
relations have small scatter ($\approx 20\%-23\%$), so we expect these relations to be relatively
unbiased.  The same cannot be said of the $L_X$--$\lambda$ relation, which exhibits
a very large scatter ($\approx 70\%$).


\subsection{Results}

Our results are summarized in Table \ref{tab:scalings}.   We again caution that we expect the $L_X$--$\lambda$
relation to be subject to large corrections from selection effects.  A proper treatment can be found
in \citet{rykoffetal12}.
We do not recommend drawing any conclusions from the $L_X$--$M$ relation quoted here; we have included
this information for completeness purposes only, and we will not discuss it further.


\begin{deluxetable*}{llllllll}
  \tablewidth{0pt}
  \tablecaption{X-ray Scaling Relations with Richness
    \label{tab:scalings}
  }
  \tablehead{
    \colhead{Catalog} &
    \colhead{Redshift Bin} &
    \colhead{$\lpivot$} &
    \colhead{Amplitude} &
    \colhead{Slope} &
    \colhead{Scatter} &
    \colhead{Eq. Mass Scatter} &
    \colhead{Outlier Fraction}
    }
\startdata
XCS & [0.1,0.3] & 40.9 & $1.129 \pm 0.056$ & $0.56 \pm 0.14$ & $0.194 \pm 0.055$ & $0.25 \pm 0.09 $  &1/25 \\
XCS & [0.3,0.5] & 49.7 & $1.283 \pm 0.071$ & $0.57 \pm 0.15 $ & $0.234 \pm 0.062$ & $0.32 \pm 0.10 $ & 0/24  \\
XCS & [0.1,0.5] & 45.6 & $1.206 \pm 0.044$ & $0.57 \pm 0.10 $ & $0.225 \pm 0.042$ & $0.30 \pm 0.07 $ & 1/49\vspace{0.04in} \\
\hline \vspace{-0.04in} \\
MCXC & [0.1,0.2] & 48.9 & $0.289 \pm 0.053$ & $1.23 \pm 0.12$ & $0.66 \pm 0.04$ & $0.32 \pm 0.03$ & 0/159 \\
MCXC & [0.2,0.3] & 65.5 & $0.927 \pm 0.086$ & $1.24 \pm 0.13$ & $0.77 \pm 0.07$ & $0.41 \pm 0.05$ & 0/85  \\
MCXC & [0.3,0.5] & 62.9 & $0.818 \pm 0.082$ & $1.57 \pm 0.11$ & $0.64 \pm 0.06$ & $0.32 \pm 0.04$ & 0/71\vspace{0.04in} \\

MCXC & [0.3,0.5] & 52.3 & $0.478 \pm 0.041$ & $1.38 \pm 0.07$ & $0.70 \pm 0.03$ & $0.36 \pm 0.02$ & 0/326\vspace{0.04in} \\
\hline \vspace{-0.04in} \\
ACCEPT & [0.1,0.5]  & 94.5 & $1.905 \pm 0.032$ & $0.407 \pm 0.066$ & $0.196\pm 0.021$ & $0.253 \pm 0.036$ & 0/54\vspace{0.04in} \\
\hline \vspace{-0.04in} \\
Mantz & [0.1,0.5] & 106.0 & $0.062\pm 0.052$ & $0.72\pm 0.12$ & $0.212 \pm 0.032$ & $0.210 \pm 0.040$ & 1/29\vspace{0.04in}
\enddata
\tablecomments{
Amplitude, slope, and scatter refer to the $X$--$\lambda$ scaling
relation, where $X$ is the relevant X-ray observable, i.e. $T_X$ for XCS and ACCEPT, $\Mgas$ for $Mantz$, and
$L_X$ for MCXC. Our convention is $\avg{\ln X|\lambda} = A + \alpha \ln (\lambda/\lpivot)$ where
$A$ is the amplitude, $\alpha$ is the slope, and $\lpivot$ is taken to be the median cluster richness.
``Eq. Mass Scatter'' refers to the scatter in mass at fixed richness estimated based on the observed
X-ray scaling relation as described in the text.  We emphasize the $L_X$--$\lambda$ relation suffers from large
systematic errors due to unmodeled selection effects.
}
\end{deluxetable*}

\begin{figure*}
  \begin{center}
      \epsscale{0.6}
    \plotone{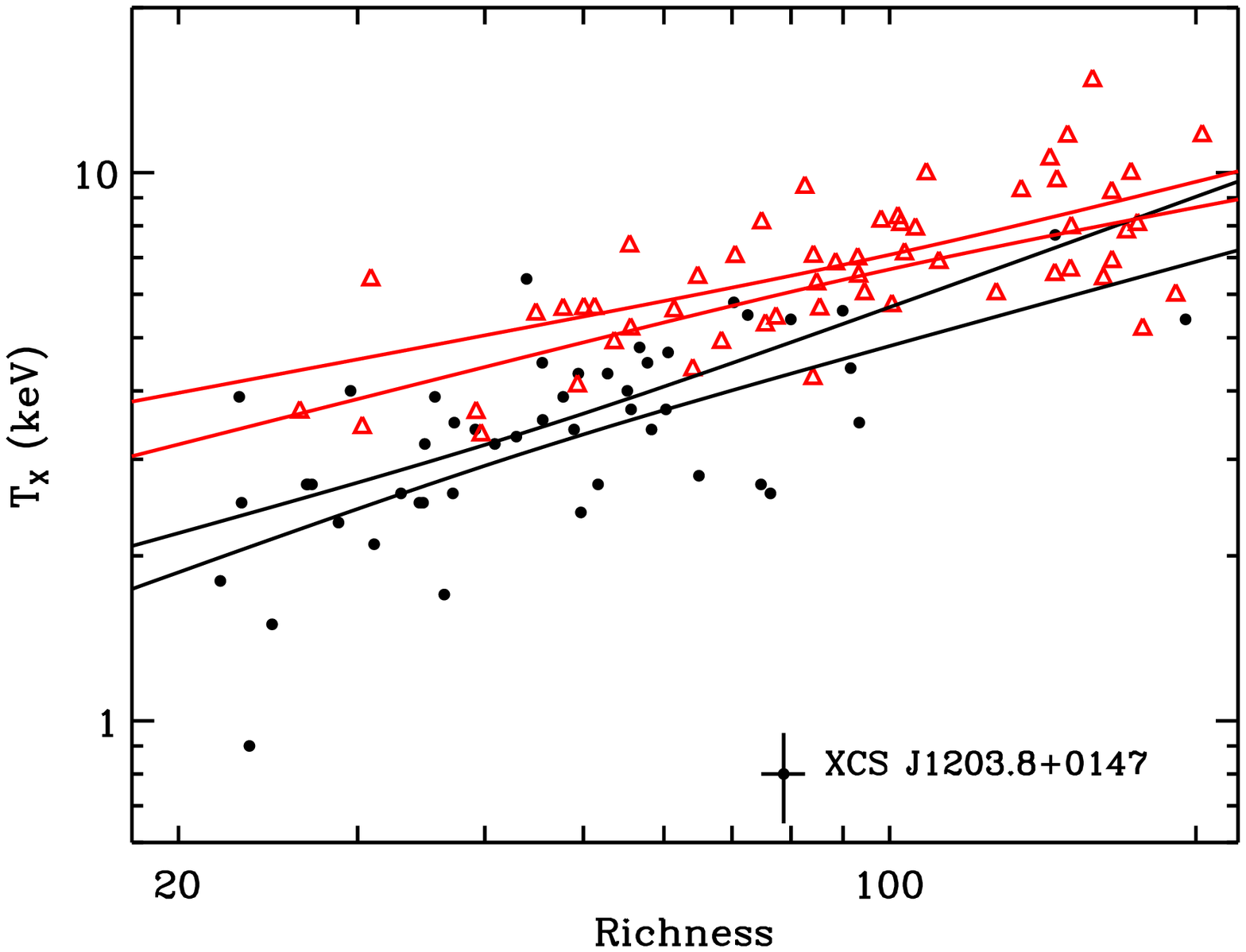} \hspace{-0.3in} \plotone{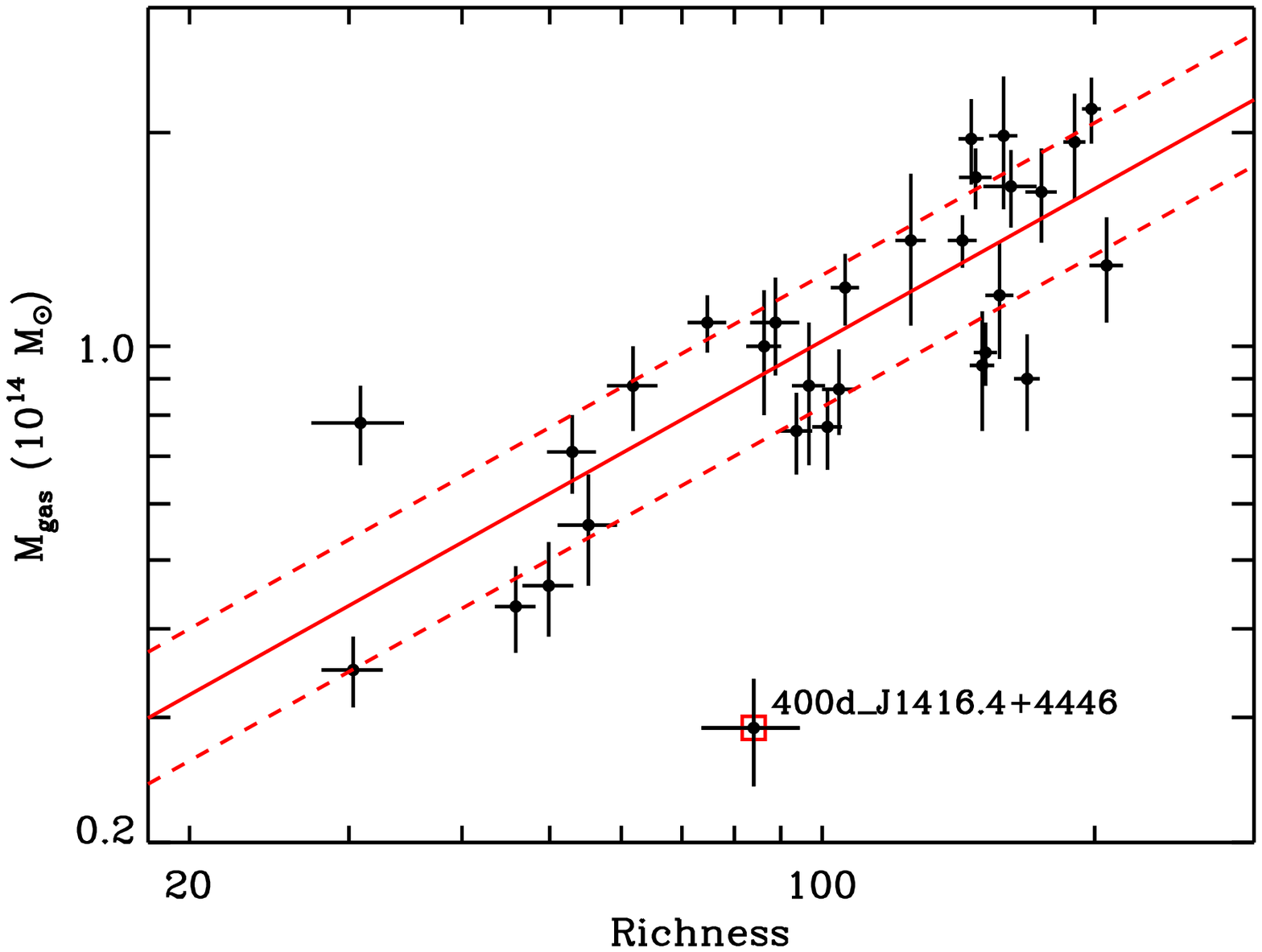}
    \caption{{\it Left panel: } The $T_X$--$\lambda$ relation for XCS (black points)
and ACCEPT (red triangles) galaxy clusters.  Errors not shown to avoid
cluttering the plot.  
The bands show the uncertainty in the mean relation.  We note the two data sets
exhibit a $\approx 40\%$ systematic offset in X-ray temperatures.  This is due
to several reasons, including the fact that the ACCEPT temperatures are
core-excised while the XCS temperatures are not.
{\it Right panel: } The $M_{gas}$--$\lambda$
relation for Mantz galaxy clusters.  The solid line shows the mean relation, and
the dashed lines shows the intrinsic scatter.  The labeled galaxy clusters
are gross outliers, and are discussed in the text. The inferred mass scatter
from this sample is  $\sigma_{\ln M|\lambda}=0.21 \pm 0.04$.}
    \label{fig:scaling}
  \end{center}
\end{figure*}

The left panel in Figure \ref{fig:scaling} shows the $T_X$--$\lambda$ relation as probed by the XCS and ACCEPT cluster catalogs.
The slopes and scatters of the two relations are consistent with each other, but their amplitudes are not.
The amplitude offset at the geometric mean of the pivot point of the two samples is
$0.35$.  That is, the two data sets exhibit a $\approx 40\%$ systematic offset in X-ray temperatures.  
This is unusually large, but there are many differences between the two X-ray temperature 
definitions, including the fact that the ACCEPT temperatures are core-excised whereas the XCS temperatures
are not.
Consequently, we are not particularly concerned with the XCS--ACCEPT offset.  Rather, 
the most significant result from this comparison is that the scatter in mass at fixed
richness inferred from the two samples are consistent with each other.  
We compute the inverse variance weighted mean of the
inferred mass scatter from the XCS and ACCEPT catalogs to arrive at
$\sigma_{\ln M|\lambda}=0.26\pm 0.03$.


We can compare these results to those derived from the $\Mgas$--$\lambda$ relation as probed by the
Mantz galaxy clusters, shown in the right panel of Figure~\ref{fig:scaling}.  The inferred mass scatter from this sample is $\sigma_{\ln M|\lambda}=0.21 \pm 0.04$,
also consistent with the XCS and ACCEPT values.  The inverse variance weighted mean from the $\Mgas$
and $T_X$ analysis is $\sigma_{\ln M|\lambda} = 0.24\pm 0.02$.  We expect selection effects
can impact the scatter by $\approx \pm 0.05$, so $\sigma_{\ln M|\lambda}=0.24 \pm 0.05$ is a more reasonable
estimate of the scatter in mass at fixed richness.  

These results are in excellent agreement with those of \citet{rykoffetal12}.  Relative to that work, our analysis
benefits from our reliance on low-scatter mass proxies, albeit at the expense of introducing systematic uncertainties
due to selection effects.   In this context, we note that while the 
inferred mass scatter from the $L_X$--$\lambda$ relation in this work appears to be larger than that inferred
from the $T_X$--$\lambda$ and $\Mgas$--$\lambda$ relations, we do not consider this problematic;
The proper analysis of the $L_X$--$\lambda$ relation in \citet{rykoffetal12} is consistent with a $\approx 25\%$ scatter
in mass at fixed richness.
We note that our estimate for the scatter is comparable to that achieved by current SZ experiments 
\citep{bensonetal13,hasselfieldetal13}.


\subsection{Richness Failures }

There are a total of two outliers over the full XCS, MCXC, ACCEPT, and MANTZ cluster samples.  
\\

\noindent {\it XCS J1203.8+0147: } This cluster is the neighbor of a much brighter X-ray foreground cluster.  A reanalysis
of the cluster by the XCS collaboration has resulted in significantly larger uncertainties in the recovered
temperature, with $\Delta T_X/T_X = 0.46$.  This new error estimates disqualifies the cluster from our sample,
and strongly suggests that deeper X-ray
data and a detailed multi-source analysis that deblends the foreground cluster is required if this system is
to be included in our analysis.
\\

\noindent {\it 400d J1416.4+4446: } This cluster was initially reported in \citet{vikhlininetal98}, along with a spectroscopic
redshift of $z=0.400$.  Visual inspection of the cluster reveals south-easterly and south-westerly extensions
of the cluster galaxies, both of which are also apparent in a recent weak lensing analysis~\citep{israeletal12}.
Based on SDSS spectra, the redshift of these components are $z=0.390$ and $z=0.373$, suggesting these
are cluster super-positions contained within a larger super-cluster.  There is also an additional more distant
southerly structure at $z=0.397$.  In short, this is a clear projection effect in the optical.
\\

It is difficult to estimate the failure rate in our richness measurements from this data. 
The outlier fractions for XCS, ACCEPT, and Mantz
galaxy clusters are 1/49, 0/54, and 1/29 respectively, though only one of these failures is due to
a \redmapper{} failure.  This suggests a $\approx 1\%$ failure rate due to projection effects in the \redmapper{}
catalog.  This failure rate is in addition to the $\approx 0.7\%$ photometric redshift failure rate.


\subsection{SZ Comparison}
\label{sec:SZComp}

We now turn to a comparison of \redmapper{} clusters to the Planck ESZ cluster
catalog.  In this comparison, we have opted not to attempt to constrain the
$\Ysz$--$\lambda$ scaling relation directly.  This is for two primary reasons. 
First, because of its broad beam, the Planck centering errors are typically
very large, which can strongly bias the richness measurements at the Planck
centers.  Second, the $\Ysz$ values derived solely from the Planck early data
are both biased and very noisy~\citep{planck11_local}.  Therefore, we leave a
detailed analysis based on SZ follow-up of individual galaxy clusters to a
future work (Greer et al, in preparation). 

\begin{figure}
  \begin{center}
      \epsscale{1.2}  
    \plotone{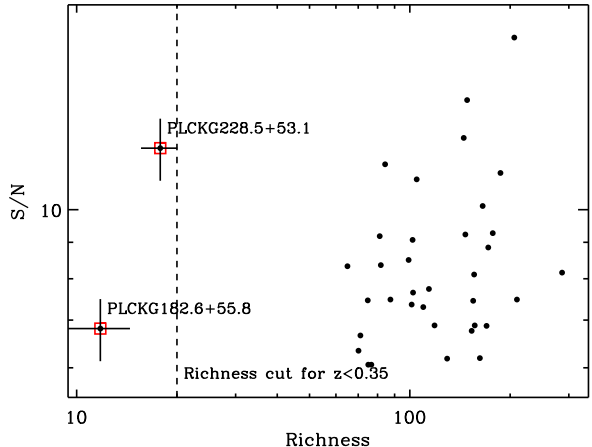}
    \caption{Signal-to-noise vs. richness for the \citet{planck11_earlysample}
cluster sample.  Richness is evaluated at the center reported in \citet{planck11_earlysample},
which can be significantly offset from the true cluster center.  The two
highlighted clusters are unusually poor and are discussed in the text. 
}
\label{fig:snlam}
\end{center}
\end{figure}

In light of these difficulties, we have opted instead to check the completeness
of the Planck ESZ cluster sample.  Figure~\ref{fig:snlam} shows the Planck
signal-to-noise as a function of cluster richness for all clusters in the DR8
footprint, evaluated at the Planck center.  Although the vast majority of the
clusters are very rich ($\lambda\gtrsim70$), there are two clusters that stand
out as unusually poor: PLCKG228.5+53.1 (ZwCl~1023.3+1257) and PLCKG182.6+55.8
(Abell~963).  Both clusters have obvious central galaxies and both are strongly
miscentered by Planck, with the radial offsets being $1.0\ \Mpc$ (6.4 arcmin)
and $0.6\ \Mpc$ (3.0 arcmin) respectively.  Both clusters are properly centered in
the \redmapper{} catalog, but even then their richnesses are relatively modest, 
$\lambda=35.5$ and $\lambda=30.5$ respectively.


Given their low richness, the detection by Planck of these two galaxy clusters
is somewhat surprising.  Visual inspection of the SDSS image of PLCKG228.5+53.1
reveals a possible very high redshift galaxy cluster just North of the quoted
Planck center ($\zphoto=0.68\pm0.04$), which suggests that this detection may in fact
be a high redshift system which was mistakenly associated with the low redshift
cluster  about $1.0~\Mpc$ away.  Deep optical follow-up of this field is desirable to either
confirm or rule out this hypothesis.

The second of the two clusters, Abell 963, is also part of the ACCEPT and
Mantz cluster catalogs.  This is a cool-core relaxed galaxy cluster, and is
also a known strong-lensing system.  Looking at the $T_X$--$\lambda$ relation
from ACCEPT, and at the $\Mgas$--$\lambda$ relation from the Mantz catalog, we
find that Abell 963 is both the hottest \emph{and} largest $\Mgas$ cluster
relative to its richness in each of these two cluster samples.  Given that the
SZ signal is proportional to $T_X \times \Mgas$, it is not surprising that
Abell 963 was detected by Planck despite its relatively low optical richness.  We have not explored whether this type of strong covariance between $\Mgas$ and $T_X$ is generic,
or whether the high thermal pressure in Abell 963 is fortuitous. Note that even though Abell 963
is the hottest and most gas-rich cluster relative to its richness, it is not an outlier, being only $2\sigma$
away from the expected $T_X$ and $\Mgas$ values.


\section{Completeness and Purity: }
\label{sec:comp_pur}

\subsection{X-ray and SZ Completeness}
\label{sec:comp}

The completeness of a cluster catalog may mean many different things depending
on the context.  As noted in Section~\ref{sec:matching}, all except one of the
reference galaxy clusters which satisfy the \redmapper{} selection threshold
are detected by \redmapper, implying that the optical incompleteness --- i.e.,
the fraction of objects that we should have detected but failed to so --- is
$\lesssim 0.3\%$.

In this section, we ask a different question: given all XCS and ACCEPT galaxy clusters with 
$T_X \geq \Tmin$, what fraction of these systems are included
in the SDSS DR8 \redmapper{} cluster catalog?
In addressing this question, we remove from consideration all redshift outliers,
both because this failure rate has already been characterized, and because
redshift errors in the reference or \redmapper{} catalogs will compromise
the relevant mass tracers ($\lambda$, $T_X$, etc.).

\begin{figure*}
  \begin{center}
      \epsscale{0.6}
    \plotone{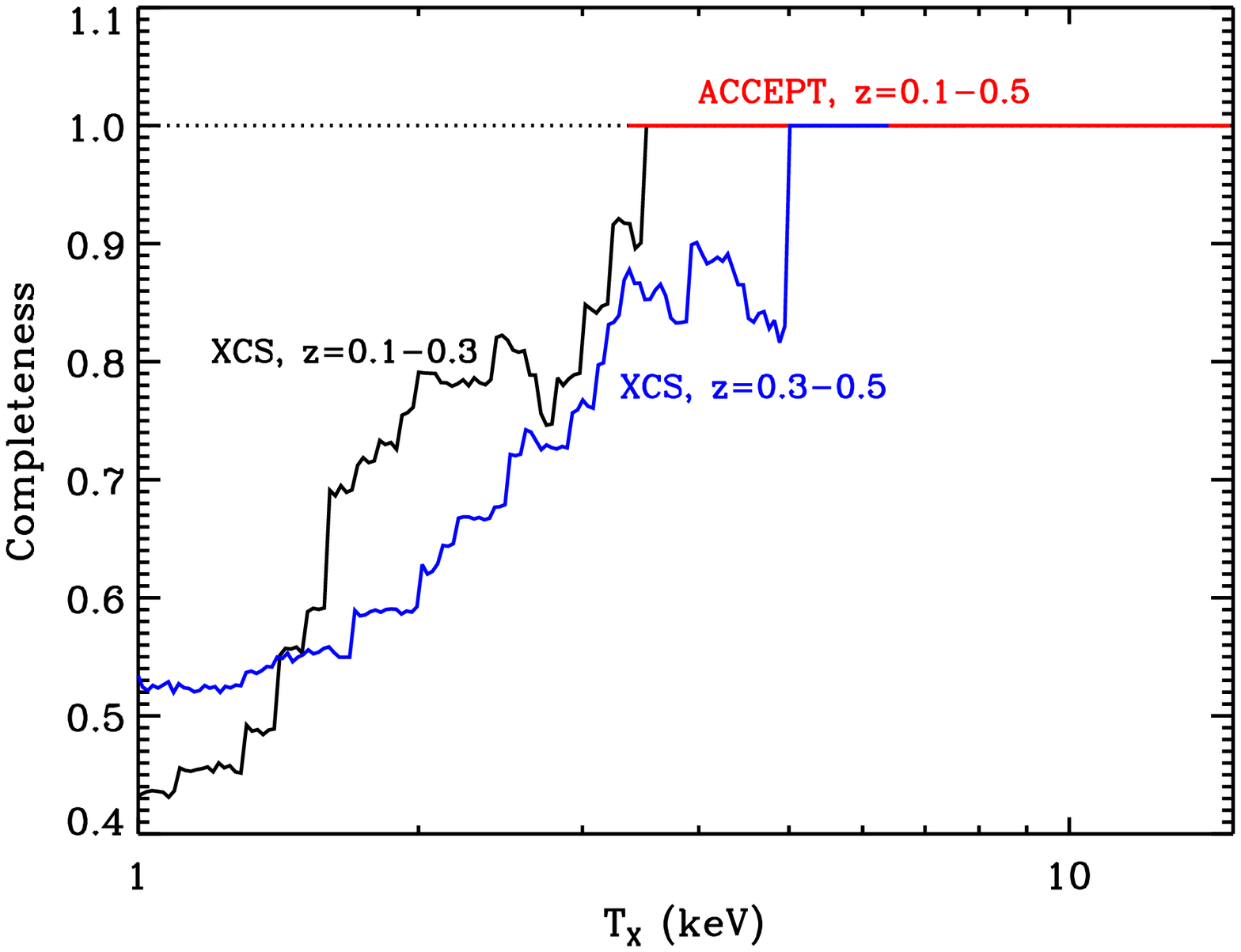} \hspace{-0.3in} \plotone{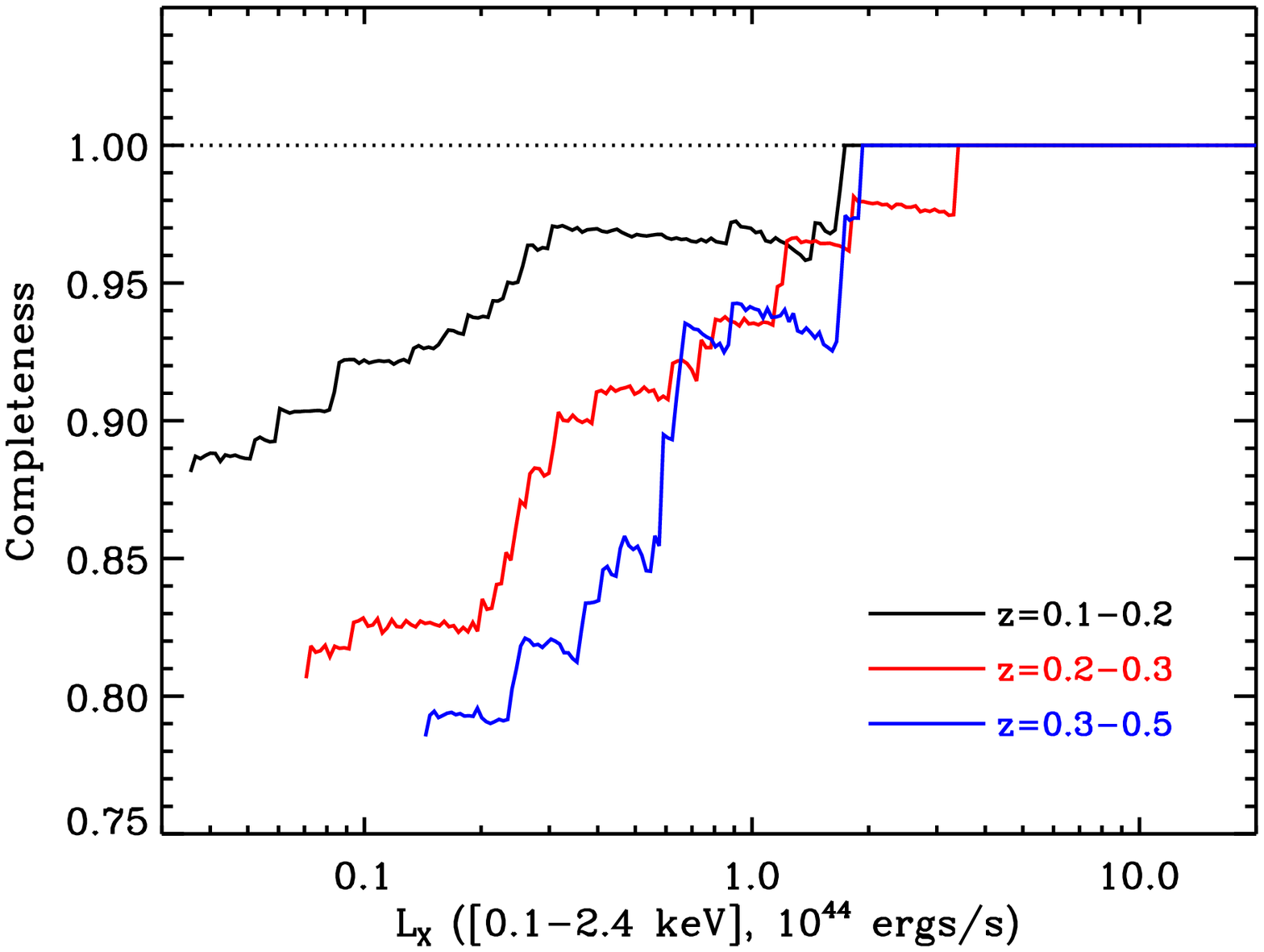}  
\caption{{\it Left panel: } Fraction of XCS or ACCEPT galaxy clusters detected in \redmapper, as a function
of the temperature threshold $\Tmin$, as labeled.  At low $z$, \redmapper{} is complete
for $T_X\gtrsim 3.5\ \keV$, with an increasing completeness threshold for $z\geq 0.35$.
{\it Right panel: } Fraction of MCXC galaxy clusters detected in \redmapper{}
as a function of the luminosity threshold $L_X$, for three different redshift bins. 
}
\label{fig:comp}
\end{center}
\end{figure*}

The left panel in Figure \ref{fig:comp} shows the fraction of XCS and ACCEPT galaxy
clusters found in the \redmapper{} catalog, as a function of the temperature threshold $\Tmin$.
As before, we limit ourselves to well measured clusters, with $\Delta T_X/T_X \geq 0.3$.  
At low redshifts, where the \redmapper{} catalog is volume limited, we detect all X-ray clusters 
with $T_X \gtrsim 3.5\ \keV$. For $z\geq 0.35$ the increasing detection threshold for \redmapper{}
systems necessarily increases the minimum temperature over which we achieve 100\% completeness.
Over the redshift bin $z\in[0.3,0.5]$, this minimum temperature is $T_X \gtrsim 5\ \keV$.


The right panel in Figure \ref{fig:comp} shows the fraction of MCXC galaxy clusters
detected by \redmapper{} as a function of the luminosity threshold $L_X$.
While the $z\in[0.1,0.2]$ and $z\in[0.3,0.5]$ bins suggest that \redmapper{} is 100\% complete above 
$L_X \gtrsim 2\times 10^{44}\,\mathrm{erg}\,\mathrm{s}^{-1}$, we find two clusters in the $z\in[0.2,0.3]$
bin above this luminosity which were not matched.
We visually inspected both of these systems:
\\

\noindent \emph{MCXC J1326.2+1230:} There is a very small ($\lambda\approx 11$) group at the reported location.
The cluster is also identified in MCXC as Abell 1735.  However,
the reported position is very distant from Abell 1735, which is easily detected by \redmapper{} ($\lambda=81.4$).
We assume the reported luminosity is appropriate for Abell 1735, in which case this cluster is in fact
properly identified in \redmapper.  
\\

\noindent \emph{MCXC J0927.1+5327:} This cluster is clearly miscentered in the optical, which has caused
the richness to fall below the detection threshold.  Thus, this cluster is less a failure to detect than it is
a case of cluster miscentering.  In Figure~\ref{fig:comp}, the drop in completeness at 
$L_X\approx 3.5\times 10^{44}\ \mbox{ergs/s}$ is due to this cluster.
\\

In light of these findings, we have matched cluster MCXC J1326.2+1230 to \redmapper{} assuming the cluster
should be at the location of Abell 1735.  The $z\in[0.2,0.3]$ curve plotted in Figure \ref{fig:comp}
includes this correction, and demonstrates that \redmapper{} is 100\% complete for 
$L_X \gtrsim 2-4\times 10^{44}\,\mathrm{erg}\,\mathrm{s}^{-1} $
clusters.

We have performed a similar analysis using the galaxy clusters in the Mantz and Planck ESZ catalogs.
We find that \redmapper{} is 100\% complete with respects to both of these catalogs.  One galaxy cluster deserves
special mention though: cluster PLCKG96.9+52.5 is clearly a rich ($\lambda=76.7$) galaxy cluster
that is not included in our SDSS \redmapper\ cluster catalog.  
Nevertheless, we do not consider this galaxy cluster a \redmapper{} failure.
When centered at the Planck reported center, the SDSS galaxy mask is such that $17\%$ of the galaxy
cluster area is masked.  However, this Planck cluster has a significant offset relative to the correct cluster
center, which is obvious upon visual inspection, and is correctly identified by the \redmapper{} algorithm.  
At that location, the mask fraction for the cluster is $\geq 20\%$,
thereby failing to pass the \redmapper{} cut.  In other words, at the correct cluster center, cluster 
PLCKG96.9+52.5 does not fall within the angular mask used to define the \redmapper{} catalog.


\subsection{Purity}
\label{sec:purity}

We now consider the converse of the question we addressed section \ref{sec:comp}:
what fraction of \redmapper{} galaxy clusters are X-ray detected?  We limit
ourselves to X-ray detections rather than SZ detections because we can probe
much lower masses by doing so.  Moreover, we have opted to use the combined ROSAT Bright
and Faint Source Catalogs as our proxy for an X-ray detected cluster.  Because these
catalogs do not require the detection of significant \emph{extended} emission
--- as is required for X-ray cluster catalogs such as NORAS --- this allows us to
probe a lower detection threshold and to improve our statistics.

We compute the fraction of \redmapper{} clusters associated with RASS sources
as a function of threshold richness $\lambda$ and redshift.  We bin the \redmapper\
clusters in narrow redshift bins of width $\Delta z = \pm0.025$, starting at
$z\in[0.1,0.15]$ and extending up to $z\in[0.45,0.5]$.  Within each redshift
bin, RASS sources are associated with a \redmapper{} cluster if the angular
separation $\Delta\theta$ is less than $800\,\kpc$ at the median redshift of
the clusters in the bin.  We have found this radius is sufficiently large to
capture most true associations while keeping the number of false matches to a
minimum.  Our chosen aperture corresponds to angles ranging from 2.0~arcmin to
5.4~arcmin depending on the redshift.   The probability of a chance association
for these apertures is  $\approx 10\%$ at $z=0.1$, falling quickly with redshift
to $\lesssim 2\%$ chance associations by redshift $z=0.35$.

\begin{figure*}
  \begin{center}
      \epsscale{0.6}
    \plotone{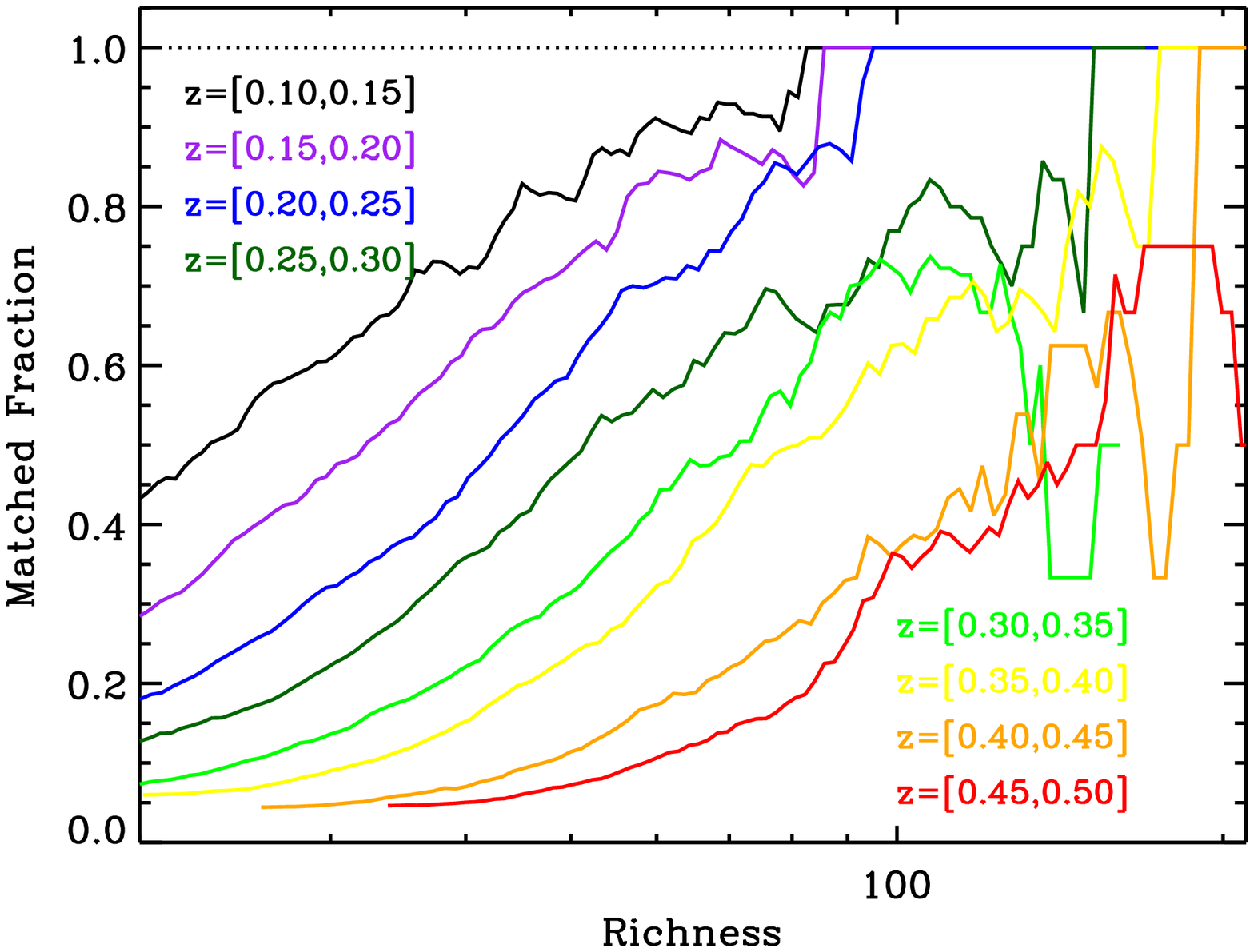} \hspace{-0.3in} \plotone{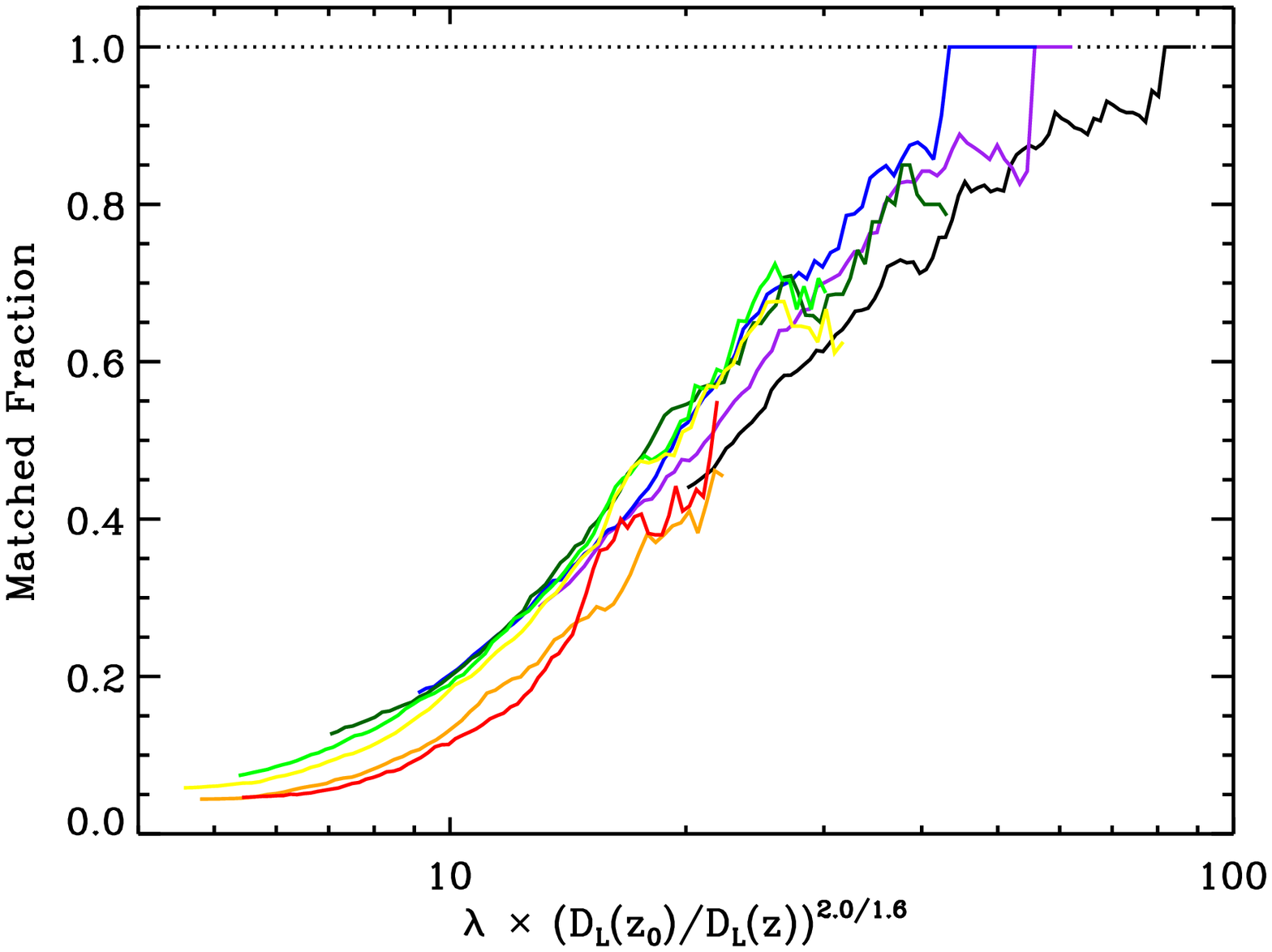}  
    \caption{{\it Left panel: } Fraction of \redmapper{} galaxy clusters above a given richness $\lambda$ matched to a source
in the combined ROSAT Bright and Faint Source Catalogs.  Each of the curves corresponds to a different redshift
bin, as labeled.  \redmapper{} clusters are matched to ROSAT sources if they fall within an angular aperture
corresponding to 800 kpc at the cluster redshift, which ranges from 2.0 arcmin at the highest
redshifts to 5.4 arcmin at the lowest redshifts.  {\it Right panel: } As left panel, but the richness is scaled by 
distance so that the $x$-axis can be considered a flux proxy.  We have ignored K-corrections and any intrinsic
evolution in the $L_X$--$\lambda$ relation.  The fact that the various curves line up demonstrates the X-ray detection
rate roughly traces a constant flux limit, as expected.
}
\label{fig:xpur}
\end{center}
\end{figure*}

Figure~\ref{fig:xpur} shows the results of this matching exercise.  As we would
expect, there is some richness $\lambdamin$ above which all \redmapper{}
clusters are X-ray detected, and this richness $\lambdamin$ increases with
redshift.  More generally, at fixed richness, the matched fraction decreases
with increasing redshift.  Note that even for our highest redshift bin,
$z\in[0.45,0.5]$, $\approx 30\%$ of the $\lambda \geq 100$ clusters are X-ray
detected.  

In our highest redshift bins, there is an obvious flattening of the
matched fraction at low richness, a clear indication that those matches are
spurious.  Interestingly, this flattening occurs at a level that is {\it higher} than
the matching rate for random points quoted above.
Thus, the flattening of the curves
appears to be impacted by the AGN rate in galaxy clusters.  Note that
the chance association rate is always much lower than unity, so the high
detection rates for rich low redshift clusters are very clearly true physical
associations.


Looking at the left panel of Figure~\ref{fig:xpur}, there is one redshift bin in particular
that looks peculiar, namely the $z\in[0.3,0.35]$
redshift bin (light green).  In this
redshift bin, the fraction of X-ray detected clusters drops at $\lambda \gtrsim
110$.  Above this richness there are nine clusters, four of which are not
detected in X-rays.  3/4 systems that are not X-ray detected exhibit 2 clear
distinct galaxy clumps, with each clump hosting a good central candidate.
In all 3 cases, the candidate central chosen by \redmapper\ is not matched
to an X-ray source, but the alternate candidate central is, suggesting
these systems are X-ray detected, but grossly miscentered.  
It is worth noting that in all 3 of these cases, \redmapper\ selected the
second candidate central as the second most likely central galaxy.
Correcting for these three systems makes the $z\in[0.3,0.35]$ curve much 
more consistent with the remaining redshift bins, but we have chosen not
to apply this correction in Figure \ref{fig:xpur}.

We now investigate whether the evolution of the curves in the left panel of
Figure~\ref{fig:xpur} is consistent with the naive expectation that the \redmapper\
galaxy clusters form a homogeneous cluster sample with a well defined $L_X$--$\lambda$
scaling relation.  Specifically, we hypothesize that the only reason why the
fraction of galaxy clusters detected in X-rays decreases is due the difficulty
of detecting low richness/high redshift clusters in RASS.  Let us then 
define a richness proxy for X-ray flux.  Assuming $L_X \propto
M^{1.6}$, and $M\propto \lambda$, we arrive at $L_X \propto \lambda^{1.6}$.
Ignoring K-corrections and intrinsic evolution in the $L_X$--$\lambda$
relation, one expect that the X-ray flux of a galaxy cluster is simply proportional to
\be
F_X \propto \lambda^{1.6}D_L(z)^{-2}
\ee
where $D_L(z)$ is the luminosity distance. Thus, a cluster of richness $\lambda$ at redshift $z$ has the same X-ray flux as a 
cluster of richness $\lambda_0$ at redshift $z_0$ where
\be
\lambda_0 = \lambda \left( \frac{D_L(z_0)}{D_L(z)} \right)^{2.0/1.6}.
\label{eq:scaled}
\ee
Assuming that X-ray detection is limited only by an effective flux threshold, it follows that all the curves in the left panel
of Figure \ref{fig:xpur} should scale onto each other if we scale the $x$-axis according to Eqn. \ref{eq:scaled}.
This is shown in the right panel of Figure \ref{fig:xpur}, where we have set $z_0=0.13$ as appropriate for our lowest
redshift bin.  The good agreement between the various curves demonstrates that the X-ray detection rate is limited
primarily by the effective flux limit of RASS.


\section{Centering}
\label{sec:centering}

One of the most difficult questions to address within the context of optical cluster finding 
concerns finding the center of galaxy clusters.
While in many cases there is a clear dominant
cD galaxy that can be adopted as the cluster center, often times there can be more than one
such candidate central.  This ambiguity can often be removed with the addition of high resolution X-ray and/or SZ 
data~\citep[e.g.,][]{menanteauetal12,songetal12,stottetal12,mahdavietal12,vonderlindenetal12}.
Roughly speaking, the correct central galaxy is the largest cD galaxy closest to the X-ray/SZ center.
We emphasize, however, that having a good centering proxy is critical for the visual selection to be robust.
We now investigate how often \redmapper{} fails to select the correct central galaxy 
by evaluating its performance in galaxy clusters with high resolution X-ray data.

We first perform visual inspection of the \redmapper{} clusters in the the XCS, ACCEPT,
and Mantz galaxy cluster catalogs to determine the correct central galaxy
for each cluster.  We restrict our analysis to
this sub-sample of galaxy clusters because of the excellent X-ray centroiding that
can be achieved with high resolution X-ray instruments.  
Ultimately, our decision of which galaxy is the correct cluster central
is subjective, but proximity to the reported X-ray center
was most often the primary criterion used to select the correct central galaxy: 
we would not attempt this analysis without the X-ray data. 
We ignore photometric redshift failures in our centering analysis, and we also remove
any clusters where the X-ray gas clearly extends beyond the X-ray
detector area, a problem that affects some of the XCS systems.
Notes on galaxy clusters that merit some discussion are collected 
in Appendix \ref{app:centering_notes}.

\begin{figure*}
  \begin{center}
        \epsscale{0.6}
    \plotone{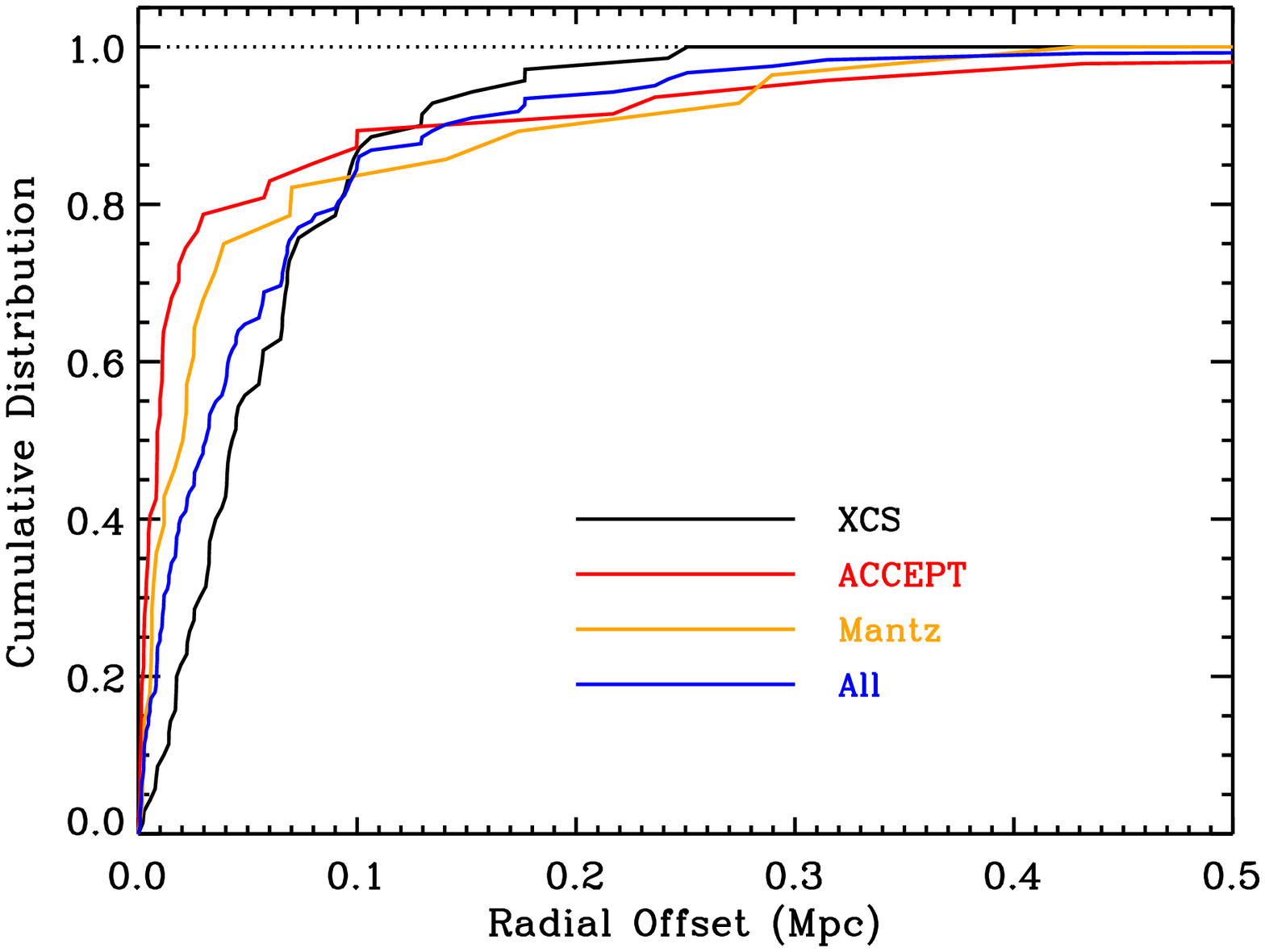} \hspace{-0.3in} \plotone{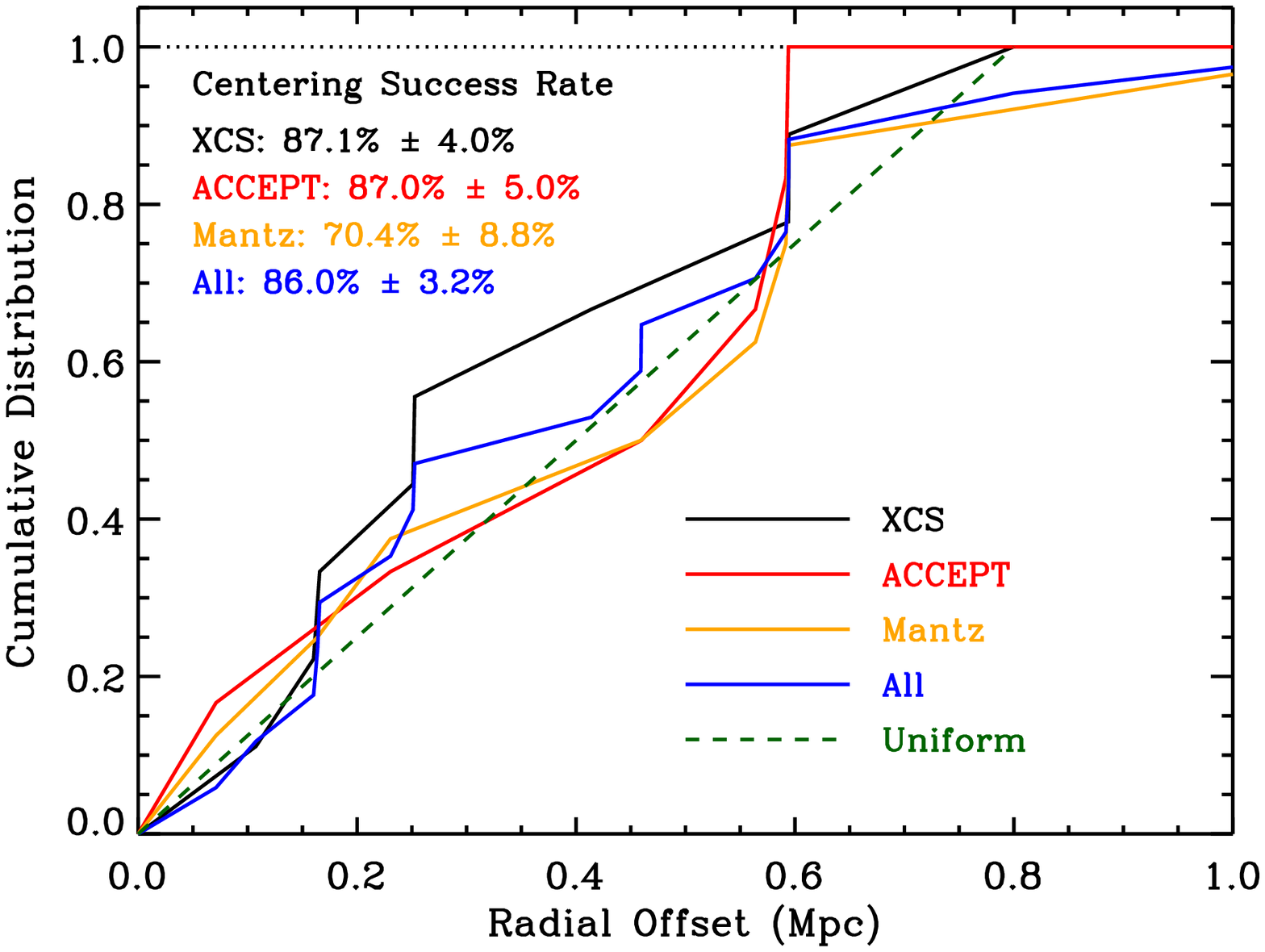}
\caption{\emph{Left Panel:} Offset distribution between the central galaxy
  selected by visual inspection and the reported X-ray center for clusters with
  high resolution X-ray imaging. \emph{Right
    Panel:} Distribution of radial offsets between our visually assigned
  cluster centers and the \redmapper{} centers for the $14\%$ of central
  galaxies that were not correctly selected.  The \redmapper{} centering success rate --- i.e., the frequency
with which our choice of central galaxy agreed with \redmapper{} --- is in the
figure legend.  
The dashed dark-green line corresponds to a uniform miscentering model that extends to $R_{\mathrm{max}}=0.8\ \Mpc$.}
\label{fig:eye}
\end{center}
\end{figure*}

The left panel in Figure~\ref{fig:eye} shows the offset distribution between
the central galaxies chosen in our visual inspection and the X-ray center
reported in each of our reference catalogs.  It is obvious that there are two
distinct cluster populations contributing to the overall distribution.  The
first population accounts for $\approx 80\%$ of the X-ray clusters, and is
comprised of X-ray clusters where the X-ray centroid and the position of the
central galaxy are essentially in perfect agreement $(R\lesssim 50\ \kpc$).
The remaining $20\%$ of the galaxy clusters are merging systems where the gas
is significantly offset from the central galaxy, though it's very rare to find systems
with an offset larger than $R\approx 300\ \kpc$.


We turn now to explore how often did \redmapper\ choose the correct central
galaxy.
We combine all three reference catalogs into a single collection of
121 unique X-ray clusters with high resolution X-ray imaging.  We find that the
\redmapper{} algorithm successfully recovers the central galaxy selected via
visual inspection $86.0\% \pm 3.2\%$ of the time.  
The right panel of Figure~\ref{fig:eye} shows the distribution of radial
offsets between our visually assigned cluster centers and the \redmapper{}
centers for the $14\%$ of central galaxies that were not correctly selected.
The figure legend also indicates the centering success rate for \redmapper{}
for each of the individual cluster catalogs.  The offset distribution can be
roughly approximated by a uniform distribution extending out to $0.8\,\Mpc$
(dashed green line).  Better statistics are required to obtain a more accurate
centering model, as there are only 17 clusters in this figure.


\section{Comparison to Other Optical Catalogs}
\label{sec:optical_comparison}

We now turn our attention to a comparison of the performance of \redmapper{} to
that of other photometric cluster finding algorithms that have been applied to
SDSS data.  The structure of this comparison follows closely the structure of
the paper as a whole.  As our goal in this section is to compare \redmapper{}
to other catalogs, as opposed to providing a detailed characterization of all
cluster catalogs, our analysis in this section is purposely less thorough than
that presented previously for the \redmapper{} catalog alone.  Specifically, we
rely solely on cylindrical matching (see Section~\ref{sec:cylindrical}), and we
do not take additional steps such as visual inspection of cluster matches to ensure 
the matchings are appropriate.   In addition, we do not take care to find all redshift
outliers between each pair of cluster catalogs.  These simplifications
necessarily have a small quantitative impact on the recovered statistics, but
still allow us to fairly compare different cluster catalogs.
In particular, in this section we recompute the \redmapper{}
statistics in exactly the same way as is done for the additional optical
catalogs, and refer the reader to the previous sections for the robust
statistics of the \redmapper{} catalog performance.

\subsection{Comparison Catalogs}
\label{sec:opt_cats}

We compare the \redmapper{} cluster catalog to four additional SDSS optical
cluster catalogs.
\\

\noindent {\bf maxBCG: } maxBCG is a red-sequence cluster finding algorithm, where cluster members
are selected on the basis of magnitude-independent color cuts~\citep{koesteretal07}.  Clusters are centered at the brightest
cluster member, and the richness is the total number of cluster galaxies above a luminosity threshold $0.4L_*$.
The catalog spans $7500\ \deg^2$,
and is limited to the redshift range $z\in[0.1,0.3]$~\citep{koesteretal07a}.
\\

\noindent {\bf gmBCG: } gmBCG is a generalization of maxBCG which does not rely on a pre-parameterized model for the
red-sequence of galaxy clusters as a function of redshift~\citep{haoetal10}.  Instead, the algorithm decomposes the color distribution of galaxies
in a given field into two Gaussian components using an error-corrected Gaussian Mixture Model (the ``gm'' in gmBCG).  In fields
where there is a galaxy cluster, one of the two gaussians in the GM decomposition is narrow.  Galaxies that contribute
to the narrow Gaussian component are identified as cluster members.  Clusters are centered at the brightest
cluster member, and the richness is the total number of cluster galaxies above a luminosity threshold $0.4L_*$.
The catalog spans $\approx 8,200\,\deg^2$ across
a redshift range $z\in [0.1,0.55]$.
\\

\noindent {\bf AMF: } The Adaptive Matched Filter cluster
catalog~\citep{spdpg11} is based on a maximum likelihood algorithm.  In this
sense, some of the philosophical underpinning of the algorithm are similar to
those of \redmapper, but there are some crucial differences.  Specifically,
unlike \redmapper, the AMF algorithm does not rely on red-sequence galaxies.
Rather, it utilizes all galaxies, relying on photometric redshift estimates to
estimate cluster membership.  In addition, the AMF centers galaxy clusters by
maximizing the cluster likelihood over position in the sky, so the center of a
galaxy cluster need not coincide with the brightest cluster galaxy.  We also
note that the AMF algorithm employs spectroscopic data where available in order
to assign cluster redshifts.  The cluster richness is the total luminosity of
the clusters in units of $L_*$.  The catalog spans $\approx 8,400\,\deg^2$
across a redshift range $z\in[0.045,0.78]$.

{\it Note: } Because AMF relies on spectroscopic data where available, we
remove clusters with spectroscopic redshifts from the sample when estimating
the photometric redshift performance of AMF.
\\

\noindent {\bf WHL: } The WHL cluster catalog~\citep{wenetal12} utilizes galaxy
photometric redshifts and a Friends-of-Friends algorithm to group galaxies into
distinct clusters.  The cluster redshift is estimated using the median
photometric cluster redshift of cluster members, and the richness is the total
cluster luminosity in units of $L_*$.  The catalog spans $\approx
14,000\ \deg^2$ across a redshift range $z\in[0.05,0.78]$.

{\it Note: } As can be seen in Appendix \ref{app:whl}, WHL employs a
spectroscopic redshift cut such that clusters with $|\zspec-\zphoto| \leq
0.055$ are discarded from the cluster catalog.  Consequently, we evaluate
the redshift performance of the WHL catalog with spectra that are
exclusive to DR9, which was not available at the time WHL was published.  In
addition, we note that WHL removed 3.6\% of their clusters after visual
inspection.  When estimating the fraction of catastrophic photometric redshift
outliers we present two results: one for the catalog as published, and the other where
we assume that half of the visually removed clusters were redshift outliers.


\subsection{Comoving Densities and Data Homogenization}

One important difficulty in comparing different cluster catalogs is that one generically expects
the performance of all cluster finders to degrade as one moves to lower
richness systems.  Consequently, a fair comparison must restrict itself to the
``same'' selection threshold.  In practice, each cluster finder uses a different richness
and/or selection threshold definition, so there is no unique way of enforcing this condition.
Here, we use the comoving density of the galaxy clusters as a proxy for selection
threshold.

\begin{figure}
  \begin{center}
  \epsscale{1.2}
    \plotone{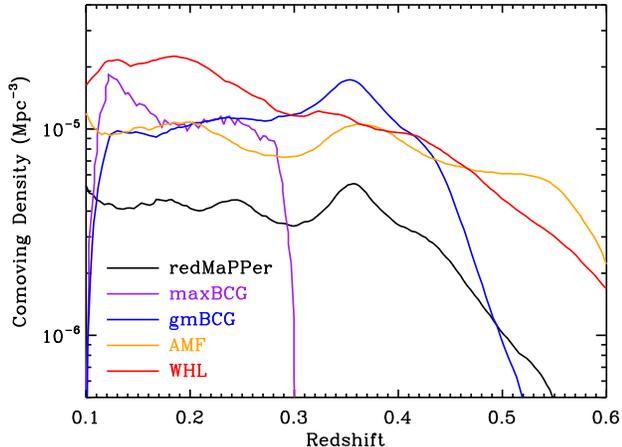}
\caption{Comoving density as a a function of redshift for galaxy clusters in each of the five optical cluster catalogs 
under consideration.  A fair comparison of galaxy clusters requires a uniform selection threshold, which we approximate
by demanding that all catalogs have the same redshift-dependent space density as the \redmapper{} catalog.  This is implemented
using a redshift dependent richness cut.
}
\label{fig:densities}
\end{center}
\end{figure}

Figure \ref{fig:densities} shows the comoving density of galaxy clusters for
each of the five optical cluster catalogs we consider.  We see that 
\redmapper{} contains fewer systems per unit volume than any other cluster
catalog.  As discussed in Paper I, this is by design: we have been purposely
conservative in our application of a selection threshold for the SDSS DR8
\redmapper{} sample in order to ensure the highest possible data
quality.\footnote{For example, were we to relax the \redmapper{} selection
  threshold to $\lambda=5$, the \redmapper\ comoving density would be
  higher than that of all other cluster catalog by more than a factor of 2.}
As the comoving density of the \redmapper{} catalog is smaller than all the
other cluster catalogs, we apply a redshift-dependent cut on the richness
appropriate for each individual catalog such that the resulting cluster
sub-samples match the \redmapper{} comoving density as a function of
redshift.


\subsection{Photometric Redshift Performance}

We evaluate the photometric redshift performance of the various catalogs by
using the spectroscopic redshift catalog from SDSS DR9~\citep{dr9}.
Specifically, we compare the photometric redshift estimate of each galaxy
cluster to the spectroscopic redshift of either the central galaxy (for
catalogs with such a galaxy) or the brightest cluster galaxy (for AMF, which
does not specify a central galaxy).

For each sample of galaxy clusters we estimate four statistics: the photometric redshift bias,
its standard deviation, its skewness, and the fraction of $4\sigma$ photometric redshift outliers.  
These statistics are estimated as a function of redshift as follows. First, we select all clusters
in a bin $z\pm 0.025$.  We compute the redshift offset $\Delta z = \zspec-\zphoto$,
and then estimate the median redshift and the median absolute
deviation of $\Delta z$.  We then select all galaxy clusters with 
\be
|\Delta z - \mbox{med}(\Delta z)| \leq 4\times 1.4826\times MAD
\ee
where $MAD$ is the median absolute deviation.
We then use the standard cumulant-based (aka $k$-statistics) estimator to
determine the mean, variance, and skewness of this cluster sub-sample.  
The scatter $\sigma_z$ is defined as the square-root of the estimated variance.  Finally, we estimate
the fraction of $4\sigma_z$ outliers based on the calculated mean and variance.
As noted in section \ref{sec:opt_cats}, when evaluating the AMF catalog we restrict our analysis to clusters with photometric redshifts only,
and the WHL analysis is restricted to clusters with spectra exclusive to DR9.

\begin{figure*}
  \begin{center}
        \epsscale{0.6}
    \plotone{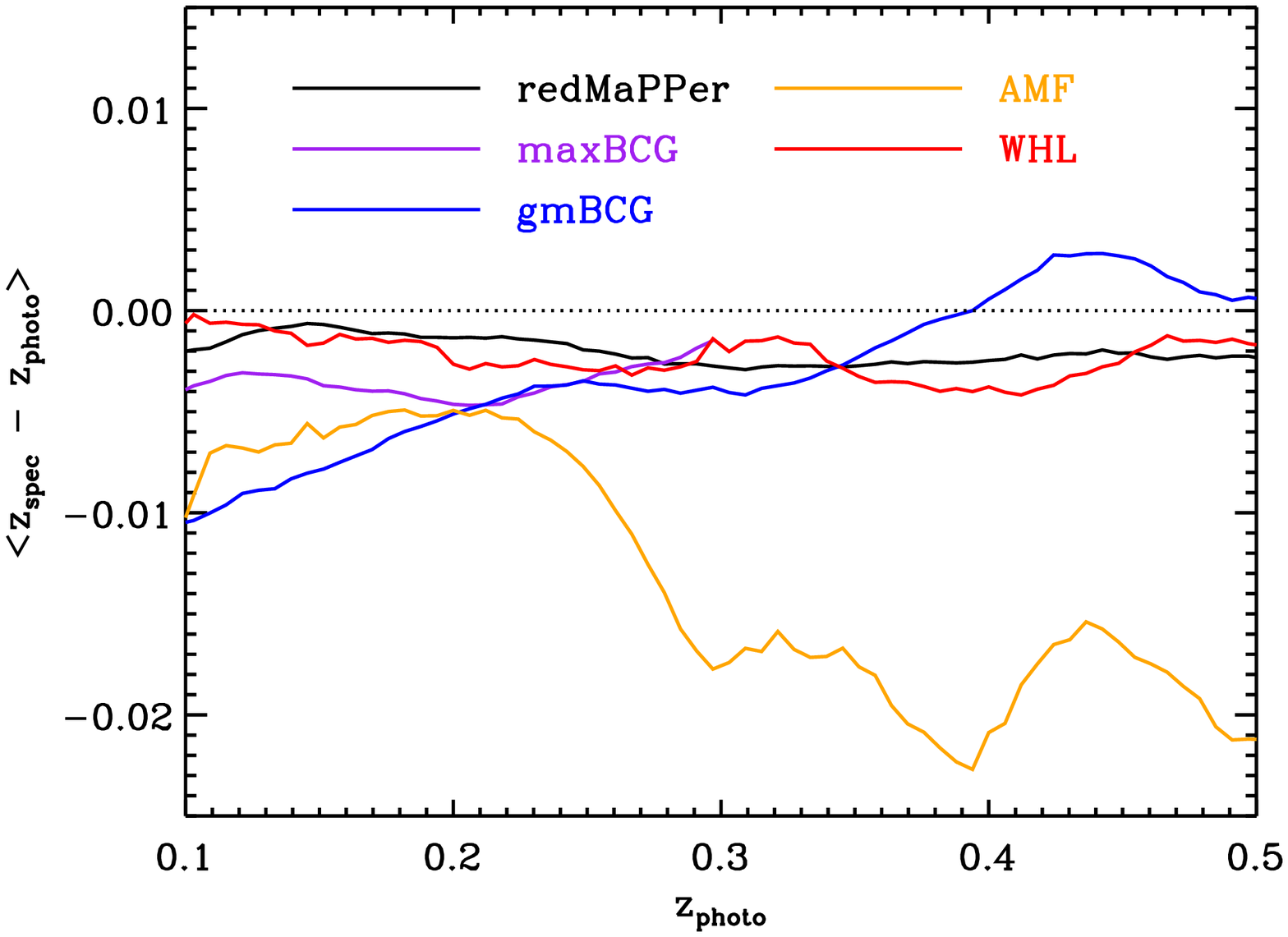} \hspace{-0.3in} \plotone{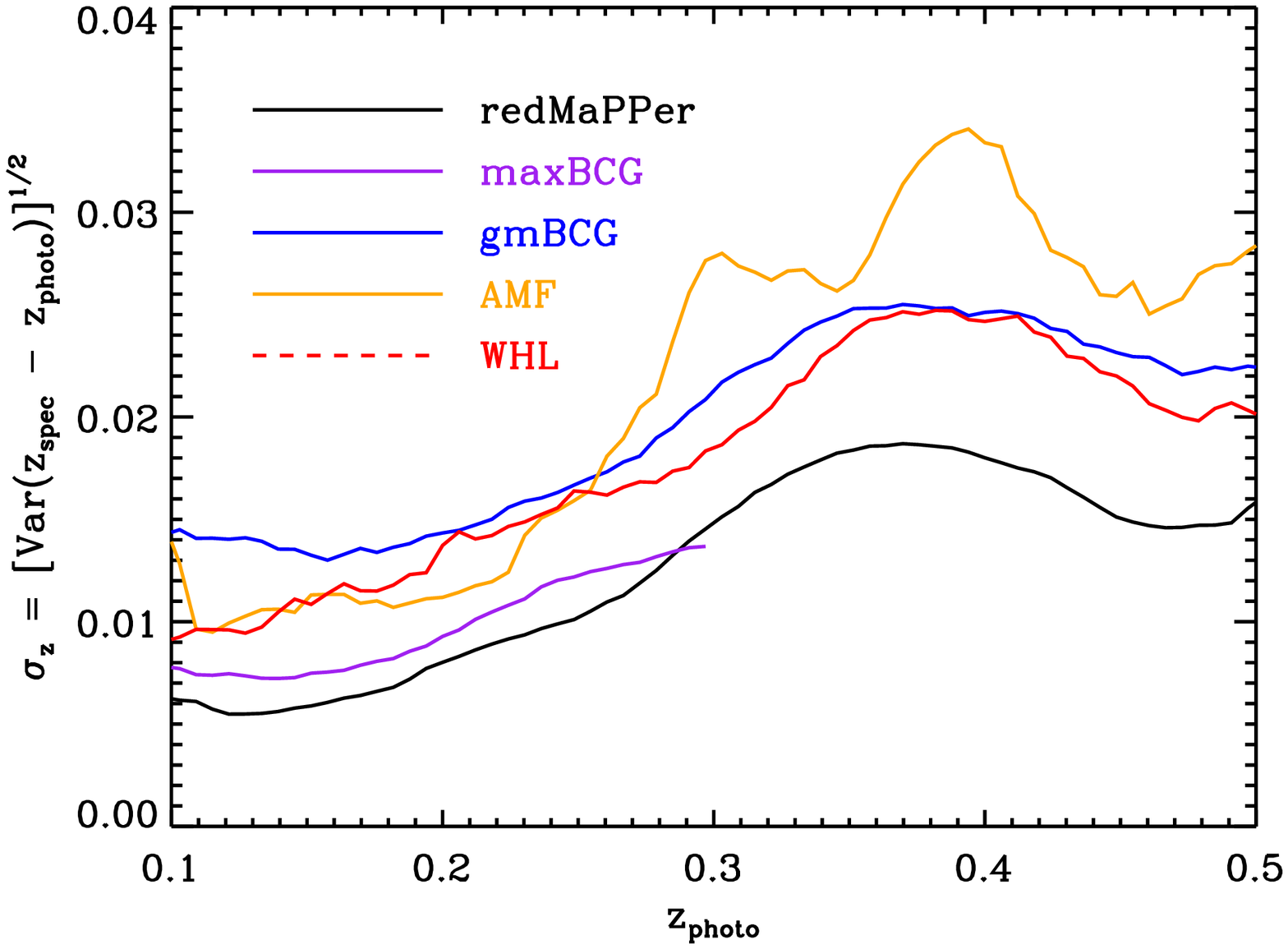}
    \plotone{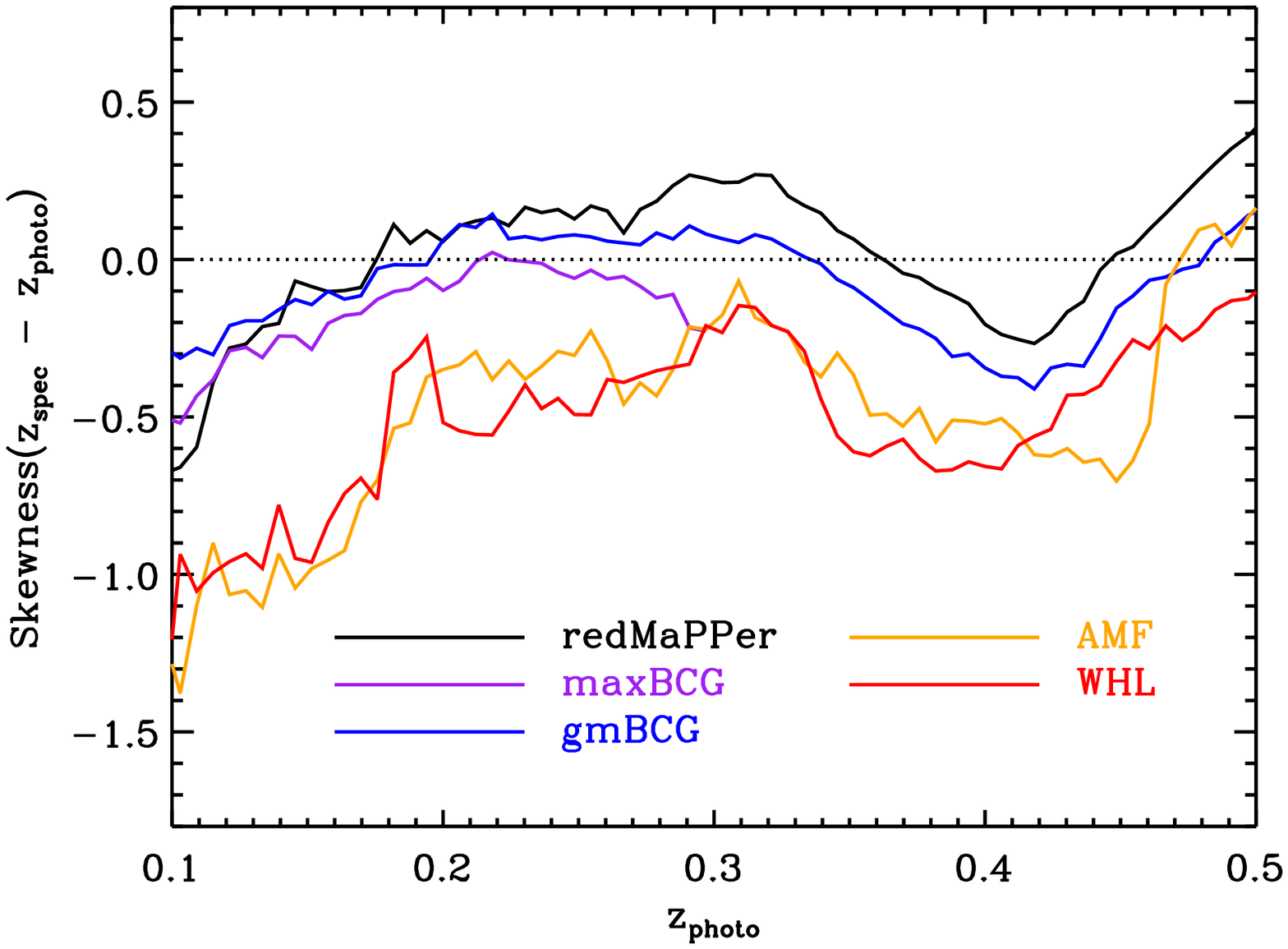} \hspace{-0.3in} \plotone{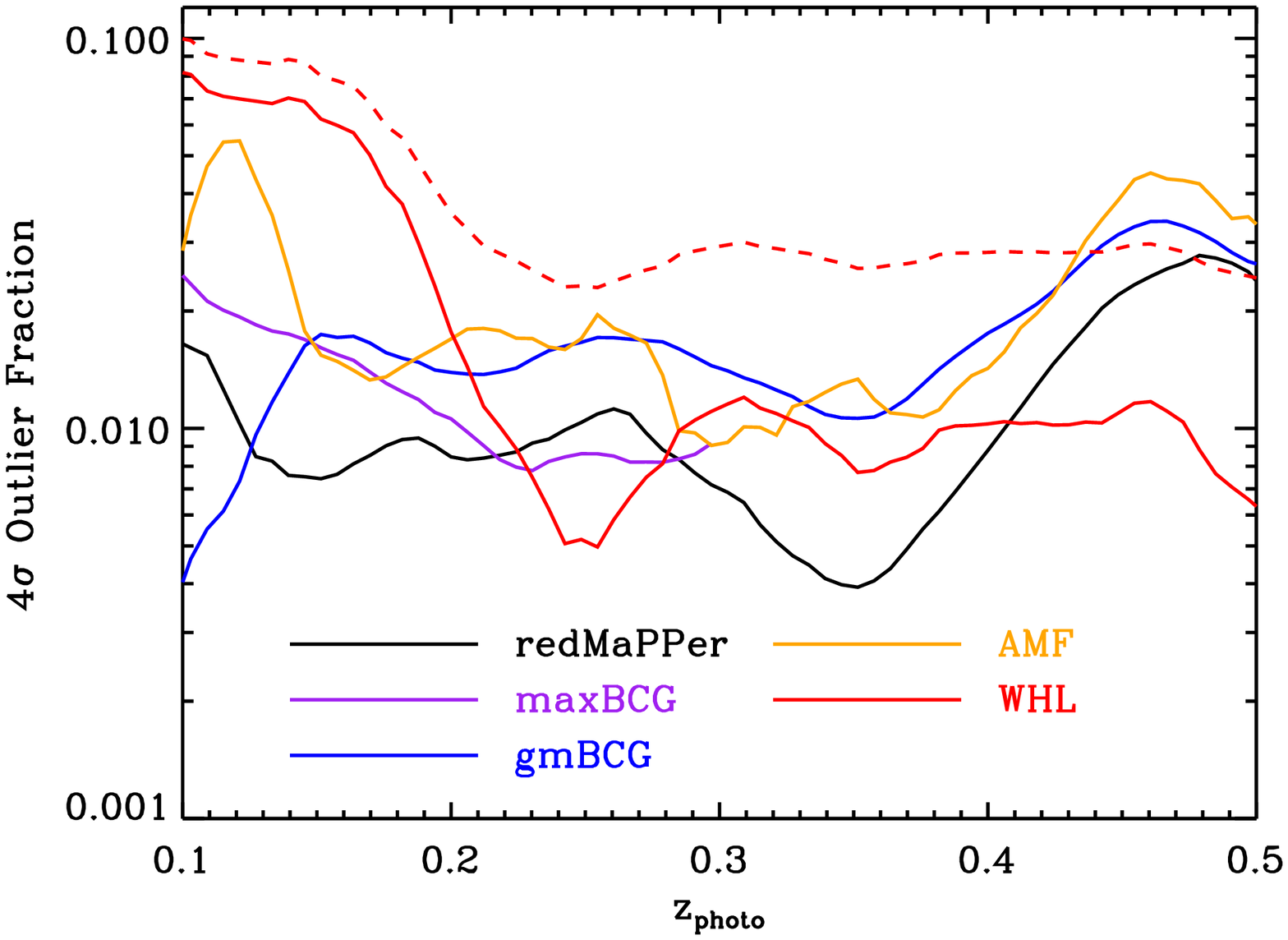}
\caption{{\it Top-Left: } Photometric redshift bias as a function of redshift for each of the five optical
catalogs under consideration. {\it Top-Right: } Photometric redshift scatter.
{\it Bottom-Left: } Skewness of the photometric redshift offset distribution. {\it Bottom-right: } Fraction
of $4\sigma$ redshift outliers.  The dashed red-line assumes half of the 3.6\% catastrophic failures
in the WHL catalog that were removed from visual inspection were redshift
outliers (see text). 
}
\label{fig:optz}
\end{center}
\end{figure*}

Figure \ref{fig:optz} collects the four statistics computed above: the photometric redshift
bias (top-left), scatter (top-right), skewness (bottom-left), and outlier fraction (bottom-right).
The least biased algorithms are \redmapper{} and WHL, while \redmapper{} and gmBCG
have the smallest skewness.  \redmapper{} outperforms the remaining
catalogs in terms of scatter and it has the lowest rate of catastrophic outliers.
In short, \redmapper{} has the best photometric redshift performance of the
five catalogs we considered.

%

\subsection{X-ray Mass Scatter}

We now compare the performance of the \redmapper{} richness estimator
to the richness estimators of other
catalogs.  To do do so, we match each optical
catalog to our three high resolution X-ray reference catalogs --- XCS, ACCEPT, and
Mantz --- using cylindrical matching with a radius of $R=1\,\Mpc$ and a redshift
width of $\Delta \zmax = 0.05$.  For simplicity, we adopt a single redshift bin of
$\zphoto\in[0.1,0.5]$.  We use the same fitting algorithm as that employed in
Section~\ref{sec:scalings} to estimate the scaling relations, including our automated $3\sigma$ outlier
rejection.

In this context, it is important to note that even if there was no correlation
between a given X-ray observable and cluster richness, our test
will always recover a finite scatter.  For instance, the scatter in 
$M_{gas}$ at fixed richness cannot
be larger than the rms in $M_{gas}$ in the Mantz cluster sample.
Consequently, in order to
measure the efficacy of a richness estimator as a mass proxy, it is not sufficient
to simply measure the scatter in X-ray observables at fixed richness; one
must also test whether the observed scatter is significantly smaller than 
that expected in the absence of any correlation.  

We estimate the significance of a sub-random scatter via Monte Carlos.
For each optical and X-ray
cluster catalog pair (e.g., \redmapper{} and XCS), we randomly scramble the
richness values among all the matched clusters and remeasure the X-ray
observable--richness scatter as above.  This procedure is repeated 1000 times
to estimate the uncertainty in the scatter for the ``random
richness'' case.  This uncertainty is added in quadrature to the error of the
recovered scatter for the unshuffled catalog, and then used
to  evaluate the significance of the difference in scatter between the
original and richness-shuffled catalogs.

As a final test the efficacy of the various cluster richness estimators, 
we have also estimated the Pearson correlation coefficient
for each of the cluster samples, limiting ourselves to clusters that are not flagged as outliers
by our automated outlier rejection algorithm.

Our results are summarized in Table~\ref{tab:opt_comparison}.  We see that in
all cases the \redmapper{} richness estimator results in the smallest scatters.
Furthermore, the reduction in scatter relative to the shuffled-richness catalog
is also most significant in the case of \redmapper\ catalog.
 In fact, \redmapper{} is the only catalog that shows
consistent evidence ($\geq 3\sigma$) that the X-ray observable--richness
scatter is significantly lower than that of a richness-shuffled catalog.
Similarly, \redmapper\ always exhibits the largest correlation
coefficients across all X-ray catalogs.  In short, the \redmapper\ richness
is clearly better correlated with X-ray temperature and $M_{gas}$ than
the richness measures of the remaining optical catalogs.

%

\begin{deluxetable*}{llllllllll}
  \tablewidth{0pt}
  \tablecaption{Comparison of the efficacy of different richness measures as
    mass tracers
    \label{tab:opt_comparison}
  }
  \tablehead{
    \colhead{} &
    \multicolumn{3}{c}{XCS} &
    \multicolumn{3}{c}{ACCEPT} &
    \multicolumn{3}{c}{Mantz}
    \\
    \colhead{Catalog} &
    \colhead{Scatter} &
    \colhead{Sig.} &
    \colhead{$r$} &
    \colhead{Scatter} &
    \colhead{Sig.} &
    \colhead{$r$} &
    \colhead{Scatter} &
    \colhead{Sig.} &
    \colhead{$r$}
    }
\startdata
\redmapper	& $0.244\pm 0.035$ & $3.0\sigma$ & 0.61 & $0.200\pm 0.021$ & $4.1\sigma$ & 0.67 & $0.201\pm 0.033$ & $6.4\sigma$  & 0.83 \\
maxBCG 		&  $0.430\pm 0.132$ & $0.7\sigma$ & 0.40 & $0.326\pm 0.122$ & $1.6\sigma$ & 0.54 & $0.363\pm 0.064$ & $1.0\sigma$ & 0.43 \\
gmBCG 		&  $0.275\pm 0.060$ & $1.5\sigma$ & 0.56  & $0.203\pm 0.024$ & $3.5\sigma$ & 0.66 & $0.357\pm 0.052$ & $2.3\sigma$ & 0.63 \\
AMF 		&  $0.273\pm 0.042$ & $1.8\sigma$ & 0.53  & $0.222 \pm 0.037$ & $1.4\sigma$ & 0.53 & $0.227 \pm 0.094$ & $2.7\sigma$ & 0.79  \\
WHL 		&  $0.305\pm 0.034$ & $2.3\sigma$ & 0.52 & $0.233 \pm 0.020$ & $1.3\sigma$ & 0.39 & $0.335 \pm 0.058$ & $3.0\sigma$ & 0.70
\enddata
\tablecomments{The XCS/ACCEPT scatter is the scatter in $T_X$
at fixed richness, while Mantz scatter is the scatter in $\Mgas$ at fixed richness.  In all cases, "Sig." is the significance of the reduction in scatter 
relative to the case where the cluster richnesses are randomly shuffled, while
$r$ is the Pearson correlation coefficient.}
\end{deluxetable*}



\subsection{Completeness}

To estimate the X-ray completeness of a given optical catalog, we must first
determine whether any given X-ray cluster falls within the footprint of the optical
cluster catalog in question.  Unfortunately, we do not have the detection masks for each
individual cluster catalog, so we must resort to an approximate method to
make this decision.
We assume that X-ray clusters that fall
within $40\,\mathrm{arcmin}$ of any optical cluster are within the optical
footprint of the corresponding cluster catalog.  
The choice of this aperture is motivated by the angular density of
\redmapper{} clusters, which corresponds to an average of $\sim1$ cluster per
circle of radius $20\,\mathrm{arcmin}$.  The motivation for setting
this angular scale based on \redmapper\ systems is motivated by the fact
that the \redmapper\ catalog is the one with the lowest density of galaxy clusters.
Of course, our method for determining which clusters fall insider or outside
the various masks is not ideal, but it does have the benefit of treating the
footprint of all the optical catalogs in the same way.

Having established which X-ray clusters fall within the footprint of a given
optical cluster catalog, we select all X-ray systems with
$z\in[0.1,0.5]$, except for when comparing with the maxBCG catalog, in which
case we restrict ourselves to the redshift range $z\in[0.1,0.3]$.
We then use the cylindrical matching method of Section~\ref{sec:cylindrical} (with
radius $1\,\Mpc$ and $\Delta \zmax = 0.05$) to obtain a set of exclusive
two-way matches.  As in other sections, when evaluating the completeness we
enforce density matching with the \redmapper{} catalog.

\begin{figure*}
  \begin{center}
        \epsscale{0.6}
    \plotone{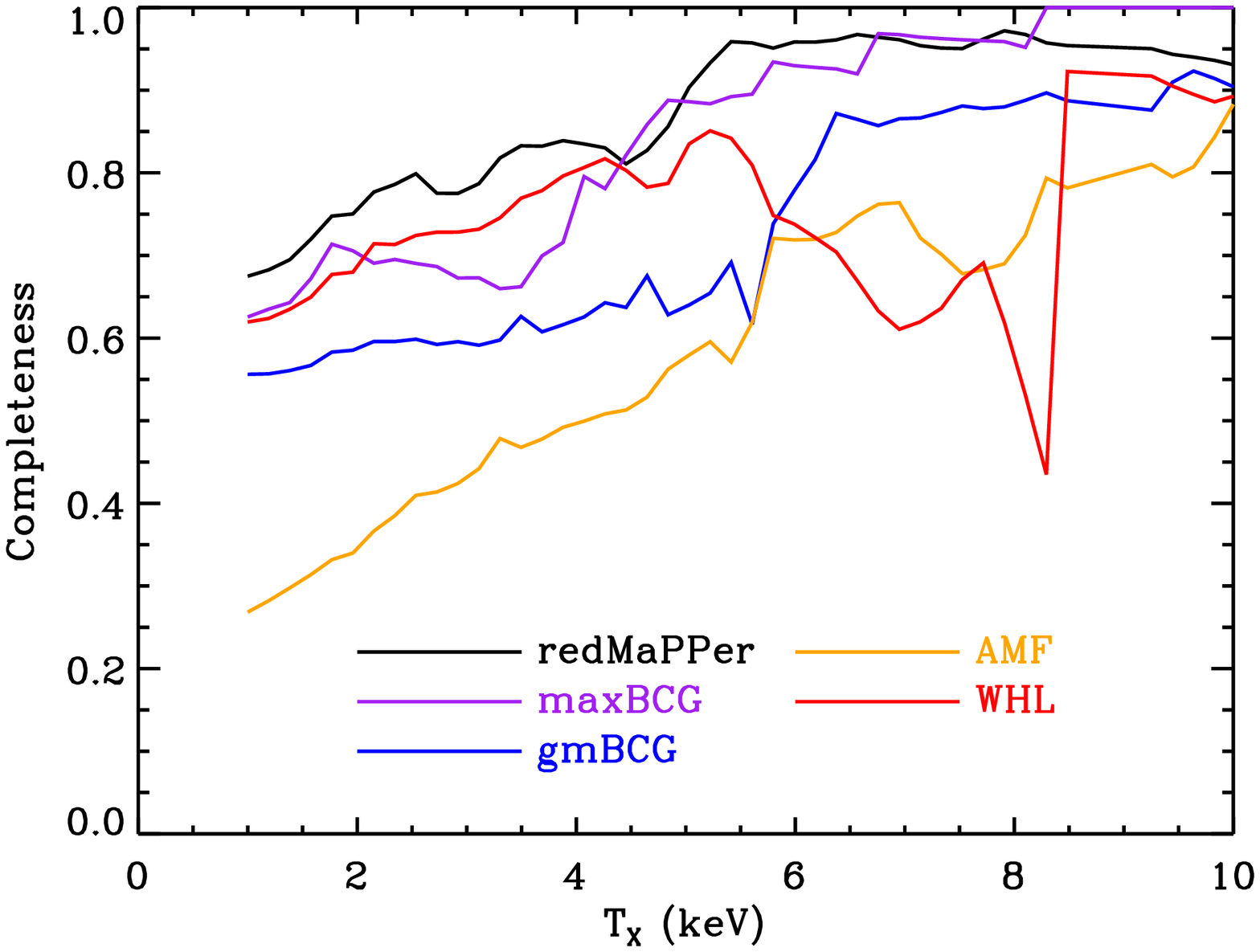} \hspace{-0.3in} \plotone{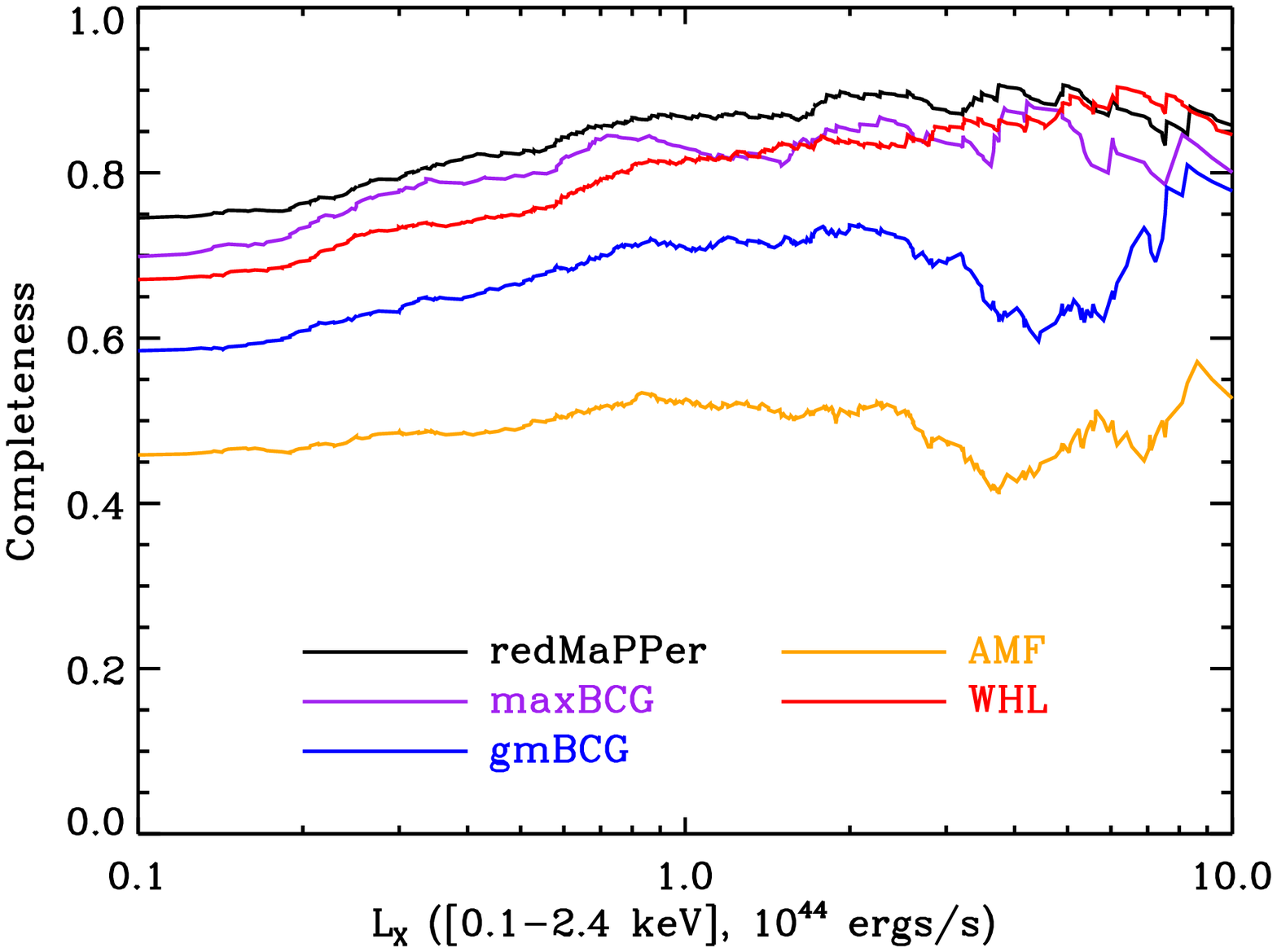}  
\caption{Fraction of X-ray clusters detected by each of the optical catalogs, as a function of
X-ray temperature (left) or X-ray luminosity (right) for all clusters with
$z\in[0.1,0.5]$ ($z\in[0.1,0.3]$ for maxBCG). 
We caution this plot uses
only an approximate treatment to determine which X-ray clusters fall within the optical masks of each
catalog, which necessarily leads to an underestimate of the completeness.  However, the treatment of the optical mask is the same
across all catalogs (see text).
}
\label{fig:opt_compare_completeness}
\end{center}
\end{figure*}

Figure~\ref{fig:opt_compare_completeness} summarizes our completeness results.
When computing the completeness as a function of $T_X$ (left panel), we compute
the completeness function using the XCS and ACCEPT catalogs, and take the
average of the two completeness functions.  When computing the completeness as
a function of $L_X$ (right panel), the luminosities are estimated using the
MCXC catalog.  We observe that in both cases the \redmapper{} catalog is the
most complete.  We emphasize that while this analysis has the virtue of
treating all optical cluster catalogs identically, the resulting completeness
functions are underestimated due to the approximate handling of the mask, the
bad redshifts in the X-ray catalogs, and the lack of optical inspection to
adjust spurious matches.  For the true \redmapper{} completeness function, see
Figure~\ref{fig:comp}.


\subsection{Purity}

Following Section~\ref{sec:purity}, we estimate the purity of each photometric
cluster catalog by matching it to the ROSAT Bright and Faint Source Catalogs.
We use four redshift bins from $z=0.1$ to $z=0.5$ in steps of $\Delta z= 0.1$.
X-ray sources are considered matched to galaxy clusters if they are within an
angular separation of $1\,\Mpc$ at the median redshift of the appropriate redshift
bin.

\begin{figure*}
  \begin{center}
        \epsscale{0.6}
    \plotone{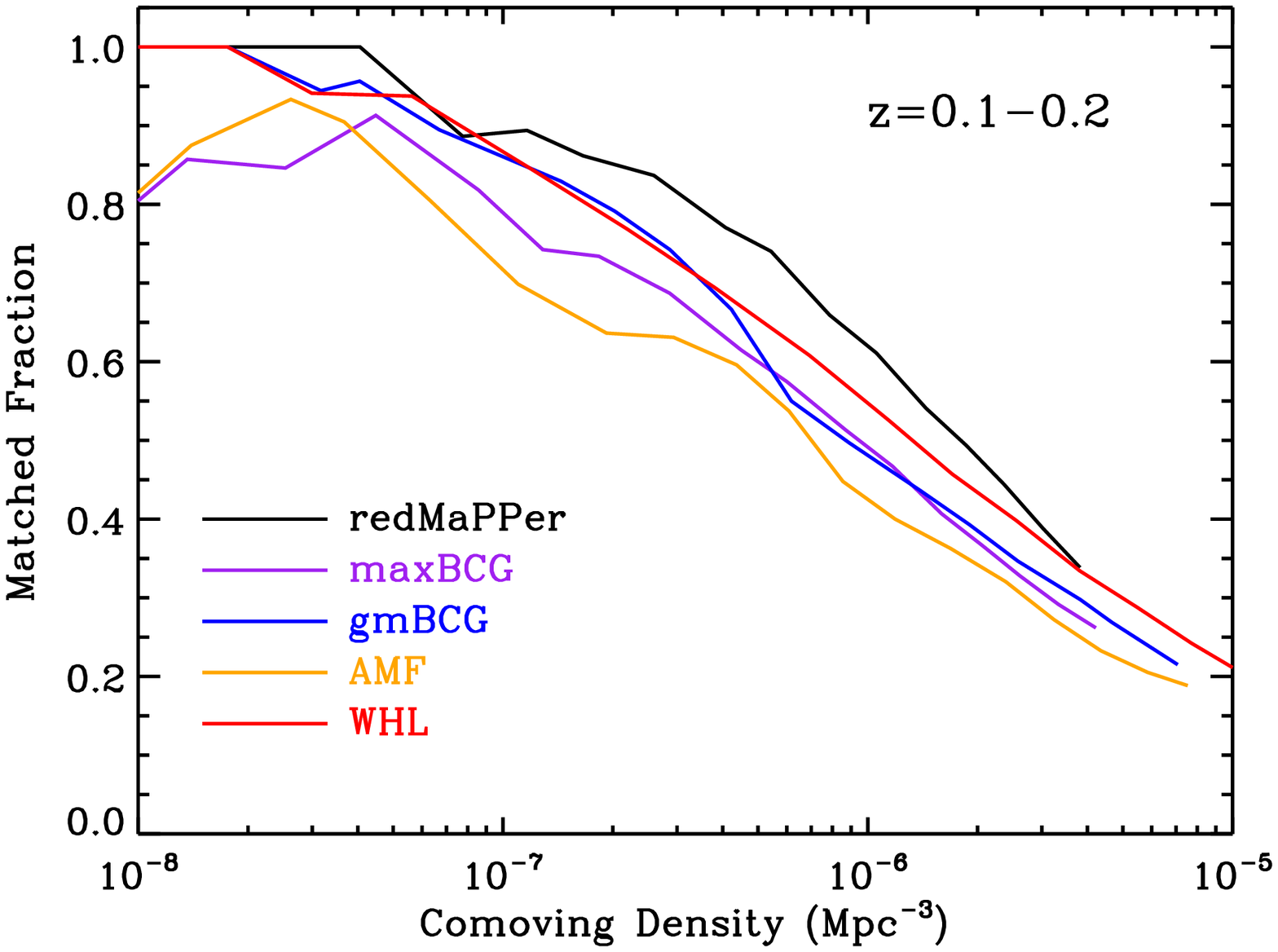} \hspace{-0.3in} \plotone{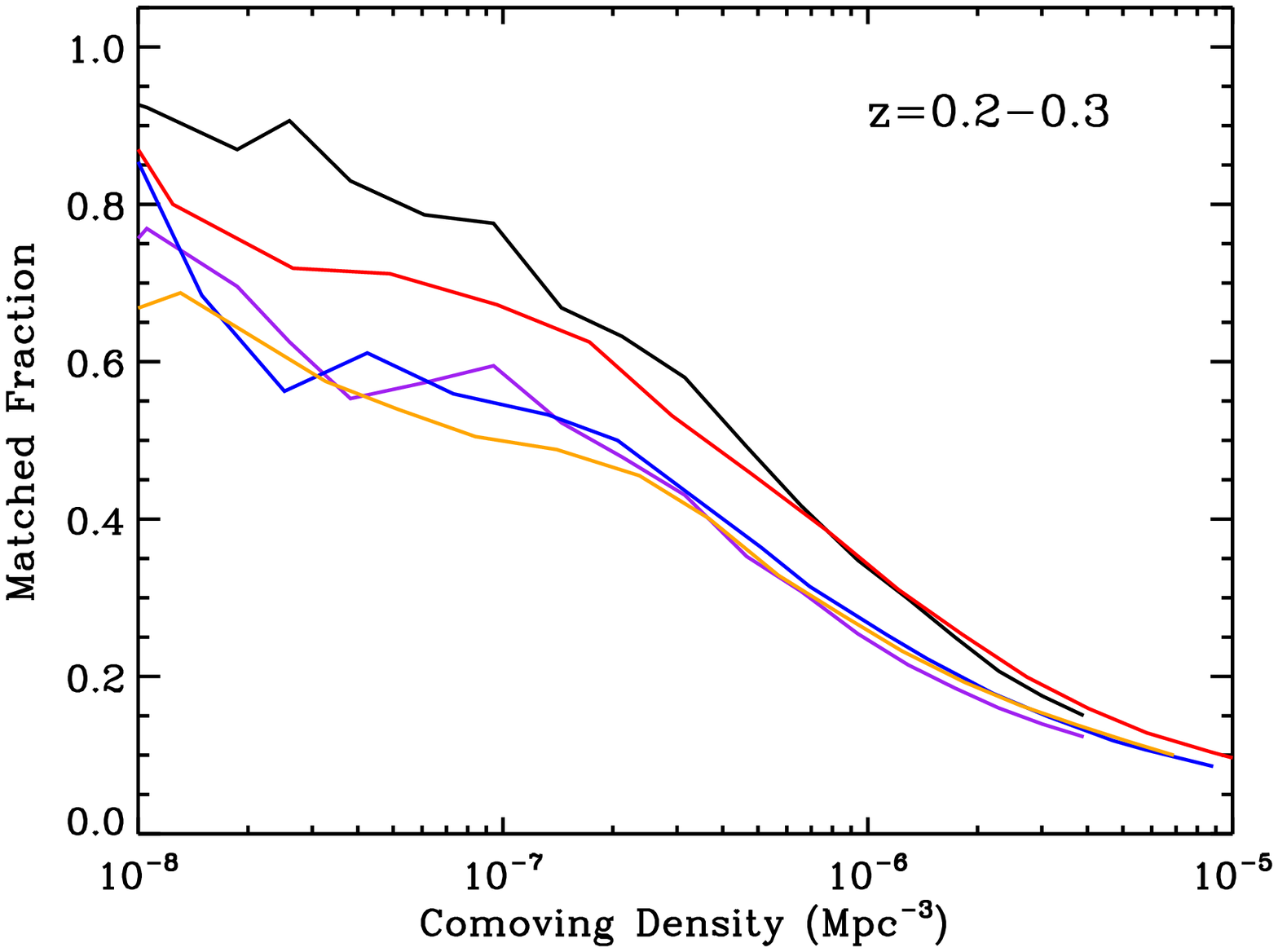}   
    \plotone{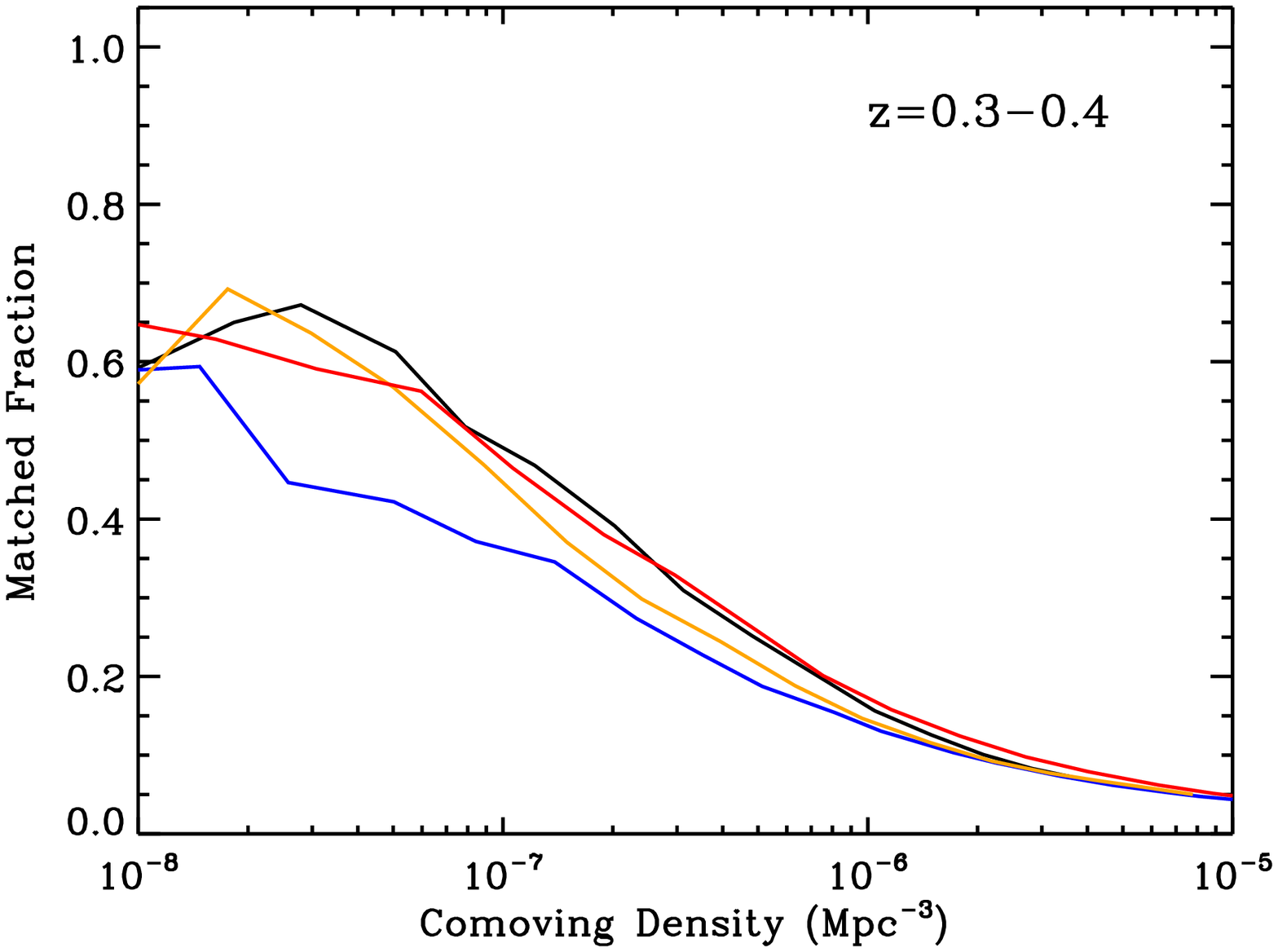} \hspace{-0.3in} \plotone{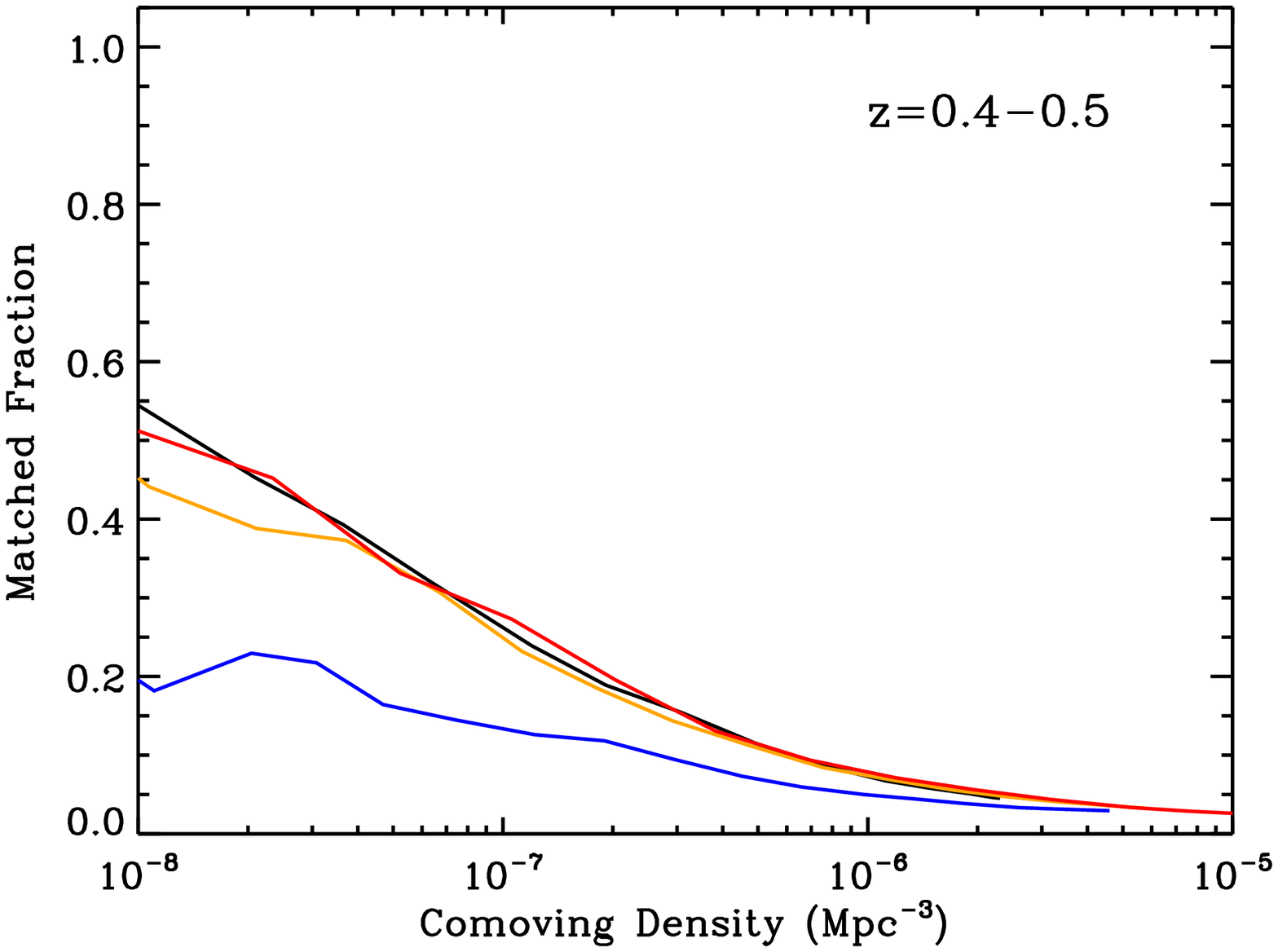}      
\caption{Fraction of optically selected galaxy clusters detected as X-ray sources in the ROSAT Bright or Faint
Source Catalogs as a function of redshift, as labeled.   
X-ray sources are matched to clusters if they are within an angular separation $\theta$
such that $\theta$ subtends $1\ \Mpc$ at the median cluster redshift within each bin.
}
\label{fig:opt_xpur}
\end{center}
\end{figure*}

Our results are summarized in Figure~\ref{fig:opt_xpur}. Because the X-ray
detection rate is a function of richness, we must properly account for the fact
that the different cluster catalogs all have different richness definitions.
Therefore, we present our results as a function of \emph{density} rather than
richness.  Each panel corresponds to a different redshift bin, and each
color corresponds to a different cluster catalog, as labeled.
We see that at low redshifts, the fraction of \redmapper\ clusters that
is X-ray detected is clearly higher than that of the remaining cluster catalogs.
By the $z\in[0.3,0.4]$ redshift bin --- the AMF and \redmapper\ algorithms
have essentially identical performance, and at $z\in[0.4,0.5]$, the \redmapper,
AMF, and WHL catalogs all have comparable X-ray detection rates.


These results clearly indicate that photometric noise and survey depth play
a critical role in the performance of cluster finding algorithms.  When the
data is such that the photometric noise is low, \redmapper\ clearly outperforms
the remaining cluster finding algorithms.  As the cluster galaxies approach
the magnitude limit of the survey, however, the difference in performance
between the various cluster catalogs becomes less pronounced.
For future surveys like the DES or LSST, the redshift at which the photometric
limit becomes comparable to the magnitude of the cluster galaxies being
selected is expected to be $z\approx 0.9$ and $z\approx 1.5$ respectively.


\subsection{Centering}

As in our centering offset analysis of the \redmapper{} catalog in Section~\ref{sec:centering}, 
we match each of the photometric catalogs under consideration to
an X-ray catalog comprised of all unique galaxy clusters in the combined XCS, ACCEPT, and Mantz cluster catalogs.
For each matched cluster, we compute the radial offset between the reported X-ray center and the reported optical 
center, from which we then compute the cumulative distribution function.

\begin{figure}
  \begin{center}
  \epsscale{1.2}
    \plotone{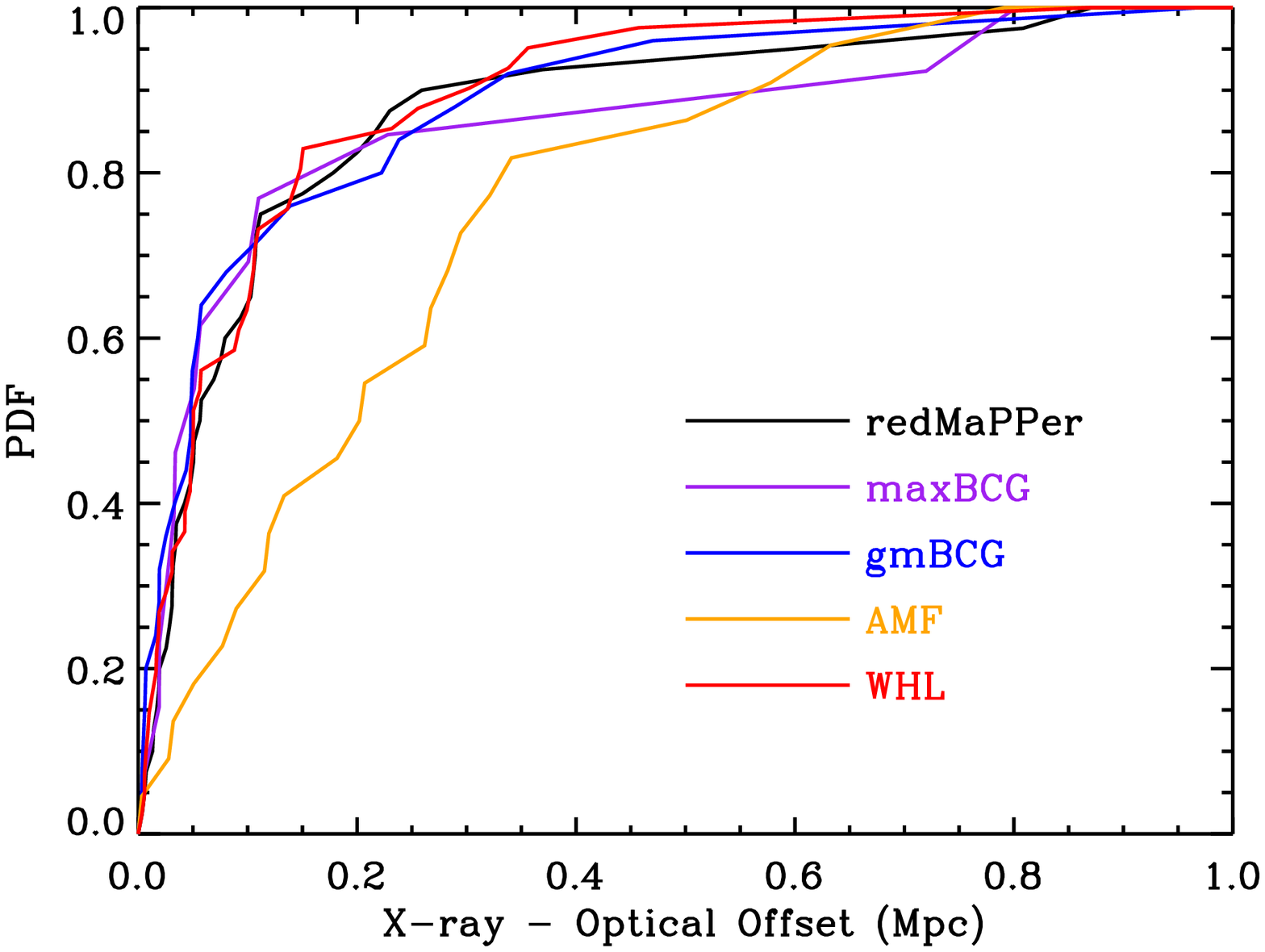}
\caption{Distribution of the radial offset between the X-ray and optical centers for matched galaxy
clusters in each of the optical catalogs under consideration, as labeled.   The X-ray
sample is comprised of all unique X-ray clusters from the combined
XCS, ACCEPT, and Mantz cluster samples.  All the catalogs have very similar
centering performance characteristics except for AMF, which is the only catalog
not to explicitly select a bright galaxy as the center.
}
\label{fig:centering_compare}
\end{center}
\end{figure}

Our results are summarized in Figure \ref{fig:centering_compare}.  
We see that the \redmapper, maxBCG, gmBCG, and
WHL cluster catalogs have essentially identical centering distributions, while AMF performs significantly
worse.  Interestingly, the AMF algorithm is the only algorithm that does not rely
on galaxy locations to define the cluster center.  Rather, their cluster centers are based on maximizing a likelihood
in the plane of the sky.  Evidently, the prior that the cluster center falls on a cluster galaxy is highly informative, and leads
to significant improvement in the centering performance.


Just as interestingly, the \redmapper, maxBCG, gmBCG, and WHL algorithms vary widely in the sophistication
of their centering algorithms, ranging from as simple as ``pick the brightest
member'', as done in WHL and maxBCG, to the iterative
self-training algorithm of \redmapper{} (see Paper I).  On the one hand, this might suggest that the \redmapper{} algorithm is unnecessarily
complicated.  On the other hand, our sophisticated algorithm has an important advantage relative to the remaining cluster finders: 
\redmapper{} is the only algorithm 
capable of estimating centering probabilities for each potential central galaxy.  That said, we note that a detailed quantitative
test of whether the \redmapper{} centering probabilities are indeed correctly estimated has not yet been performed, 
a problem that we will return in a future work.  For now, we will simply state that our internal tests have produced good evidence
that the centering probability is well correlated with the likelihood of cluster being correctly centered, i.e. clusters with low centering
probability are more often incorrectly centered.  However, whether the centering probabilities are quantitatively correct remains
to be determined.


\section{Summary and Conclusions: }
\label{sec:summary}

We have performed an extensive quality test of the performance of the SDSS \redmapper{} cluster 
catalog by comparing it against several X-ray and SZ cluster catalogs.  We find that
the \redmapper{} redshifts are nearly unbiased, with biases falling at the $0.005$ level or below.
In addition, our photometric redshifts exhibit remarkably low scatter, as low as $\approx 0.006$
at low redshift, but increasing with redshift due to the increase in photometric noise in the data.
Moreover, the scatter in redshift is nearly symmetrical, and has a very low ($\approx 1\%$)
catastrophic failure rate.  In this context, it is important to emphasize that the word ``catastrophic''
here means that the true cluster redshift is more than $4\sigma$ away from our photometric redshift
estimate.  Because our \photoz{} scatter is so low, clusters with $|\zspec-\zphoto|$ as low as $\approx 0.025$
can be considered catastrophic failures.
We note too that in addition to providing a simple photometric
cluster redshift, the \redmapper{} catalog includes an estimate of the full redshift probability distribution
$P(z)$ for every galaxy cluster (see Paper I).

We have also tested the performance of our richness estimator $\lambda$ \citep[see also][]{rozoetal09b,rykoffetal12}
using our reference X-ray catalogs.  In particular, using the XCS and ACCEPT
cluster catalogs we have estimated the
scatter in X-ray temperature at fixed richness, and converted it to an equivalent scatter in mass at fixed richness.
Both data sets are consistent with each other, and the combined analysis suggest a scatter in mass at fixed richness
of $26\%$.   We have further verified this analysis by utilizing the Mantz cluster sample to measure the
scatter in the $\Mgas$--$\lambda$ relation, from which we are able to recover an equivalent mass scatter of
$21\%$.  Given the systematic uncertainties associated with our measurements --- most importantly the
fact that the selection function of the clusters with combined optical and X-ray data is not well constrained ---
we estimate that a reasonable range for the scatter in mass at fixed richness for the \redmapper{} clusters
is $25\% \pm 5\%$.  Interestingly, this scatter may well be comparable to the estimated scatter from SZ mass
proxies in current SZ surveys such as ACT and SPT \citep{bensonetal13,hasselfieldetal13},
and better than what has been achieved with Planck so far using SZ data only \citep{planck11_local}.
We hasten to add, however, that we do expect significant improvement in future
Planck data releases.  

In this context, it is also worth emphasizing that we did find one clear cluster outlier in our analysis ---
cluster 400D J1416.4+4446 --- whose richness was grossly over-estimated due to projection effects.
The existence of this one outlier suggest that $\approx 1\%$ of our galaxy clusters suffer from projection
effects.  This failure rate is somewhat lower than what was estimated in Paper I, where we found that up to
$\approx 5\%$ of the galaxy clusters can be severely affected by projection effects, particularly at the rich end.
The lower failure rate measured in this work is likely to be related to our reliance on external X-ray catalog
for the majority of our tests, as X-ray selected sub-samples of the \redmapper\ clusters are expected to
be more robust to projection effects.

We have also explored the completeness of the \redmapper{} cluster catalog as a function of
X-ray luminosity, temperature, gas mass, and SZ detectability.   \redmapper{} detects all
galaxy clusters in the Mantz sample, ACCEPT sample, and Planck ESZ sample.  At low ($z\lesssim 0.35$)
redshifts, where the \redmapper{} catalog is volume limited, the catalog is 100\% complete
at $T_X \gtrsim 3.5\ \keV$, and above a luminosity threshold of $L_X \gtrsim 2\times 10^{44}\ \mbox{ergs/s}$.
Because of the large scatter ($\approx 60\%-70\%$) in the relation between X-ray luminosity and cluster
richness, the decrease in completeness with decreasing luminosity is very modest; at low redshifts,
the catalog remains $\approx 90\%$ complete all the way down to $L_X \approx 10^{43}\ \mbox{ergs/s}$.

In addition, we have estimated the fraction of \redmapper{} galaxy clusters that are X-ray detected by cross-matching
our catalog with the ROSAT Bright and Faint Source Catalogs.  As expected, all rich, low redshift
clusters are X-ray detected, with the detection fraction decreases with increasing redshift and/or
decreasing richness.  We have verified that this decrease in the detection fraction is primarily --- and most 
likely exclusively --- driven by the decrease in the expected X-ray flux of the galaxy clusters with decreasing
richness and increasing redshift.  That is, we see no evidence of a large population of intrinsically X-ray dark clusters.  
Of course, as noted above, we do have clear evidence for a population of galaxy clusters that suffers from projection 
effects, comprising $\approx 1\%$ of the cluster sample.

We have further measured the miscentering rate of the \redmapper{} galaxy clusters by visually inspecting
121 galaxy clusters with high resolution X-ray data.  Based on our visual inspection and the reported X-ray
centers, we identified the correct central galaxy for each of these clusters, and compared it to the galaxy
that was selected by the \redmapper{} cluster catalog, finding that $\approx 86\%$ of the \redmapper{}
galaxy cluster are correctly centered.   

Finally, we also performed a ``quick-and-dirty'' version of all of the above analyses on all 
photometric cluster catalogs in the SDSS published to date.  Our goal here is not to characterize in detail
the performance of each of these cluster catalogs, but rather to provide a meaningful comparison among them.
Note in particular that when performing these comparisons, the \redmapper{} catalog is treated
in the exact same way as the remaining cluster catalogs.

As the result of this comparison, we find that the \redmapper{} catalog has a higher completeness
than that of any other cluster catalog.  As for purity, we find that at low redshift the \redmapper\ algorithm
has a higher X-ray detection rate than that of any other catalog.  However, as the survey depth becomes
comparable to the magnitude of the bulk of the galaxy population of galaxy clusters, the differences between
the various algorithms decrease.  At $z\in[0.3,0.4]$, the \redmapper\ and AMF algorithm have comparable
X-ray detection rates, and at $z\in[0.4,0.5]$, the WHL algorithm also reaches an X-ray detection
rate comparable to that of \redmapper\ and AMF.  
We note that in the Dark Energy
Survey, the transition point at which the survey depth becomes comparable to the magnitude of the cluster galaxies
is $z\approx 0.9$.

Turning to the redshift performance comparison, the \redmapper{} photometric redshifts
are the least biased, with the WHL redshifts --- themselves based on the DR8 photometric redshifts ---
being equally unbiased.   The \redmapper{} redshifts clearly exhibit the lowest scatter amongst all catalogs, and, along with
the gmBCG redshift,  they exhibit the least amount of skewness.  The catastrophic
failure rate of the \redmapper{} redshifts is the lowest amongst the SDSS clusters catalogs.

We also compared the performance of the \redmapper{} richness estimator to that
of the various other cluster catalogs, both via the $T_X$--richness scaling
relations, and via the $\Mgas$--richness scaling relations.  The \redmapper{}
catalog consistently exhibited the lowest scatter on all relations.  Moreover,
the \redmapper\ catalog is the only one to consistently show evidence ($\geq 3\sigma)$ 
for the recovered scatter being lower than that estimated when the richnesses
are randomly shuffled amongst all clusters.

Finally, turning to cluster centering, we found that the centering algorithm in \redmapper{} performs
as well as --- i.e., neither better nor worse --- the remaining catalogs.  The only exception was the
AMF catalog,  which performed worse.  Since AMF is the only catalog that did not choose to center the
galaxy clusters by identifying a central galaxy, it is clear that the prior that galaxy clusters should be centered
on bright cluster galaxies is highly informative, and improves the overall performance of a cluster finder.
We also emphasize that while the remaining cluster catalogs all had equivalent centering performances, 
the \redmapper\ catalog is unique in that the algorithm estimates the probability that the chosen central galaxy is correct, and also provides
4 additional candidate central galaxies along with their corresponding probabilities.  Our hope is that cluster miscentering
can be robustly treated using these estimated probabilities, but we postpone the development of a statistical framework 
that exploits this information to a future work.   For now, we simply note that \redmapper{}
is the only algorithm for which such a framework can be developed.

As an important caveat to our results, we wish to emphasize that the results of our optical cluster comparison presented in this work 
reflect the reality of the various cluster finding algorithms as implemented in the SDSS, with the data that was available at the
time.  In particular, it is worth keeping in mind that the performance of photometric redshift cluster finders such as AMF
and WHL will depend not only on the cluster finding algorithm itself, but also on the quality of the redshifts employed 
in the construction of the \photoz{} estimates.  For instance, AMF relied on SDSS DR6 data with \photoz{} estimates from
\citet{oyaizuetal08}, while WHL relied on SDSS DR8 \photozs.  Given that there has been a great deal of improvement
between the DR6 \photozs{} from \citet{oyaizuetal08} and the DR8 \photozs{} --- in large part due to increasingly large
training samples for \photoz{} machine learning methods, with current samples reaching over 850,000 
galaxies\footnote{http://www.sdss3.org/dr9/algorithms/photo-z.php} ---  it is reasonable to expect the performance of AMF to improve
if run using DR8 \photozs.  That said, we believe that the fact that the \redmapper\ performance is in no way limited by
the performance of galaxy \photoz{} measurements is still clearly a valuable feature.  Moreover, while \redmapper\ does require
spectroscopic training samples, just as \photoz{} measurements do, it is important to emphasize that, as demonstrated
in Paper I, \redmapper\ can deliver photometric redshift quality using a spectroscopic training sample comprised
of only the $400$ brightest cluster galaxies in SDSS (sampled over the full redshift range, see paper I for details).

The excellent performance of \redmapper{} in the Sloan Digital Sky Survey provides strong evidence that upcoming photometric
surveys like the Dark Energy Survey and LSST will be capable of producing very large, high quality cluster samples.
Moreover, the excellent quality of the photometric redshifts, the low rate of catastrophic redshift failures and projection
effects, and the low scatter in the optical mass proxy, all suggest that the quality of photometric cluster catalogs should
easily suffice from the point of view of precision cosmology.  In particular, a $\approx 1\%-2\%$ catastrophic failure
rates on these algorithms is comparable to the expected precision of the cluster mass estimates that can be achieved 
in surveys like the Dark Energy Survey \citep[e.g.][]{oguritakada10,weinbergetal12}, suggesting 
that optical detection systematics are not likely to be a dominant
source of errors in near-future photometric surveys.  The most important exception remains cluster centering: with only
85\% of the \redmapper{} galaxy clusters being correctly centered, there is a clear need for a robust statistical framework
with which to address this problem, a question that we will return to in future work.


\acknowledgements

The authors would like to thank M. Donahue for her help with the ACCEPT cluster
catalog, and Alexis Finoguenov and Marguerite Pierre for useful comments on an earlier
draft of this paper.  This work was supported in part by the U.S. Department of Energy contract to 
SLAC no. DE-AC02-76SF00515. 

Funding for SDSS-III has been provided by the Alfred P. Sloan Foundation, the Participating Institutions, the National Science Foundation, and the U.S. Department of Energy Office of Science. The SDSS-III web site is http://www.sdss3.org/.

SDSS-III is managed by the Astrophysical Research Consortium for the
Participating Institutions of the SDSS-III Collaboration including the
University of Arizona, the Brazilian Participation Group, Brookhaven National
Laboratory, University of Cambridge, Carnegie Mellon University, University of
Florida, the French Participation Group, the German Participation Group,
Harvard University, the Instituto de Astrofisica de Canarias, the Michigan
State/Notre Dame/JINA Participation Group, Johns Hopkins University, Lawrence
Berkeley National Laboratory, Max Planck Institute for Astrophysics, Max Planck
Institute for Extraterrestrial Physics, New Mexico State University, New York
University, Ohio State University, Pennsylvania State University, University of
Portsmouth, Princeton University, the Spanish Participation Group, University
of Tokyo, University of Utah, Vanderbilt University, University of Virginia,
University of Washington, and Yale University.

\bibliography{mybib.bib}
\bibliographystyle{apj}

\appendix


\section{Clusters where Matchings Differed}
\label{app:vismatching}

Below is the complete list of galaxy clusters where the cylindrical and membership matching algorithms
differed.
\\

\noindent {\bf XCS:}

\noindent {\it XCS J1310.9+5720: } No cylindrical match.  
The are two clear galaxy clumps at this
redshift, with the XCS cluster corresponding to the northern component.
In \redmapper{}, the southern component is richer, so it is this southern component
that was matched to the XCS cluster.  However, the northern component is also
detected, and is the obvious correct match.

\noindent {\it XCS J0920.8+3028: } Cylindrical and membership matches differ.
This cluster is at the eastern component of Abell 781.  \redmapper{} identifies
both the eastern and western component, but the membership matching
matches the XCS system to the western component because it is richer.

\noindent {\it XCS J0943.5+1639: } Cylindrical and membership matches differ.
The membership match is correct.
The cylindrical match for this clusters is a foreground cluster
at $z=0.18$ (compared to $z=0.25$).

\noindent {\it XCS J2337.9+2711: } No cylindrical match.  Membership
match is clearly correct.

\noindent {\it XCS J0943.9+1641: } No cylindrical match.  Membership map
is a larger cluster due NW which is not included in the XCS
catalog.  The correct match is found by \redmapper{} algorithm,
but with a richness that falls below our selection threshold in Paper I.

\noindent {\it XCS J1256.8+2548: } No cylindrical match.
Membership match incorrect: correct match is a \redmapper{}
cluster that falls below our selection threshold.

\noindent{\it XCS J0840.7+3830: } No membership match.
Correct match is clearly a \redmapper{} cluster that falls below 
our selection threshold.
\\

\noindent {\bf MCXC:}

\noindent {\it MCXC J1326.2+1230: } No cylindrical match.   This
is a small satellite cluster of the much larger system Abell 1735, and does
not pass the \redmapper{} selection threshold.

\noindent {\it MCXC J1230.7+3439: } This is an interesting system.  The X-ray location
is close to a cD galaxy, which falls roughly in the middle of 2 \redmapper{} clusters
at the same redshift.  The least rich of these 2 \redmapper{} clusters is subject to masking
from the richer system, which is the membership match of the X-ray system.  In this
sense, the membership match is clearly correct, but it may be subject to miscentering.
Unambiguous evidence for miscentering would require high resolution X-ray imaging.

\noindent {\it MCXC J1227.1+1951: } No cylindrical match.  Membership match is 
clearly correct.

\noindent {\it MCXC J0751.4+1730: } No cylindrical match.  The photometry
around this region is clearly compromised.  We set this cluster to unmatched
to reflect the photometric failure, even though technically a cluster was found.

\noindent {\it MCXC J1415.2-0030: } No cylindrical match.  This cluster
is Abell 1882, which has two clear components.  \redmapper{} is centered
on the NW component, whereas the MCXC system corresponds to the SE
component.  The SE component appears to be the most massive one
based on the X-ray data, suggesting the cluster is catastrophically miscentered
in \redmapper.  The membership matching association is clearly correct.

\noindent {\it MCXC J1311.8+3227: } No cylindrical match. 
Membership match is incorrect, with correct match falling below the 
\redmapper{} selection threshold.

\noindent {\it MCXC J2258.1+2055: } Matchings differ.  Membership match is clearly correct.

\noindent {\it MCXC J0943.5+1640: } Matchings differ.  Membership match is clearly correct.

\noindent {\it MCXC J1235.1+4117: } No cylindrical match.  Membership match is clearly correct.

\noindent {\it MCXC J1337.8+3854: } No membership matching.
X-ray clusters has no nearby bright galaxies, which is very unusual.  We assume
this cluster is unmatched, and note this may be an X-ray false detection.

\noindent {\it MCXC J1254.8+255: } No membership match.  Correct match
falls below \redmapper{} selection threshold, so cluster should be unmatched.

\noindent {\it MCXC J0943.7+1644: } Matchings differ.  Membership match
is clearly correct.

\noindent {\it MCXC J1436.9+5507: } No cylindrical match.  Membership
match is a larger foreground cluster.  Cluster does not pass \redmapper{}
selection threshold, so it should be unmatched.

\noindent {\it MCXC J1254.6+2545: } No cylindrical match.  Membership match is clearly
correct.
\\

There are no clusters where the matchings differed in the ACCEPT, Mantz, and Planck reference cluster
catalogs.


\section{Notes on Unmatched Clusters}
\label{app:unmatched}

Below is the complete list of unmatched clusters in our reference catalogs.  
\\

\noindent {\bf XCS:}

\noindent {\it XCS J0943.9+1641: } The cluster was assigned a high richness because
it neighbors a richer system that is not included in XCS.  The cluster should be unmatched.
\\

\noindent {\bf MCXC:}

\noindent {\it MCXC J0751.4+1730: } This cluster was identified as a catastrophic photometric failure,
and was therefore unmatched by hand in our visual inspection.

\noindent {\it MCXC J0159.3+0030: } This is a very rich cluster, but gets masked out from \redmapper{} because 
at its optical center, more than 20\% of the cluster is lost to the galaxy mask.  In other words, the cluster doesn't
formally fall within the angular \redmapper{} selection region.

\noindent {\it MCXC J0159.3+0030: } Like the previous cluster, this system is lost to the galaxy mask since at 
its optical center, more than 20\% of the cluster is lost to the galaxy mask.  Thus,
the cluster is formally not within the angular \redmapper{} selection region.

There are also 4 clusters that are matched to \redmapper{} systems, but fall below our selection threshold
at the \redmapper{} center/redshift.  These are MCXC J0927.1+5327, MCXC J1340.9+3958, J1011.0+5339,
and J1334.5+3756.
\\

\noindent {\bf Planck ESZ:}

\noindent {\it PLCKG96.9+52.5: } Like the previous cluster, this system is lost to the galaxy mask since at 
its optical center, more than 20\% of the cluster is lost to the galaxy mask.  Thus,
the cluster is formally not within the angular \redmapper{} selection region.
\\

There are no unmatched clusters in the ACCEPT and Mantz reference catalogs.


\section{Notes on Redshift Outliers}
\label{app:zoutliers}

Below is the complete list of redshift outliers between all reference cluster catalogs.
\\

\noindent {\bf XCS Outliers:}

\noindent {\it XMMXCS J0921.2+3701: } Cluster is miscentered in optical based on X-ray data.  
The \redmapper{} cluster center has a spectroscopic redshift $\zspec=0.235$, in agreement with the 
\redmapper{} photometric redshift.   It seems likely that the reference redshift is incorrect, but a spectroscopic
redshift of the correct central galaxy is unavailable.  
\\


\noindent {\bf MCXC Outliers:}

\noindent {\it MCXC J1621.0+2546: } SDSS spectra confirm \redmapper{} redshift.

\noindent {\it MCXC J1421.6+3717: } SDSS spectra confirm \redmapper{} redshift.

\noindent {\it MCXC J1621.0+2546: } SDSS spectra confirm \redmapper{} redshift.

\noindent {\it MCXC J1017.5+5934: } SDSS spectra confirms \redmapper{} redshift.

\noindent {\it MCXC J0935.4+0729: } SDSS spectra confirm \redmapper{} redshift.

\noindent {\it MCXC J2135.2+0125: } SDSS spectra unavailable, but visual inspection strongly suggests reference redshift is incorrect.

\noindent {\it MCXC J0809.6+2811: } \redmapper{} redshift compromised by bad SDSS photometry from a nearby star.

\noindent {\it MCXC J0826.1+2625: } Our photometric redshifts at the \redmapper{} center and the reference cluster center disagree.
The cluster is properly centered by \redmapper, and the corresponding photometric redshift is correct based on that galaxy's spectroscopic
redshift.

\noindent {\it MCXC J0847.1+3449: } Optical inspection reveals the \redmapper{}  match is in fact a foreground cluster in the vicinity
of the X-ray cluster.  The correct cluster match is detected by \redmapper{}, but falls below the richness threshold, likely due
to masking by the foreground cluster.  This cluster should be formally unmatched.  We update our matchings appropriately,
and remove this system from the list of matched clusters.

\noindent {\it MCXC J1011.4+5450: } The X-ray center falls between two galaxy clumps.  Both clumps are the same photometric
redshift $z_\lambda \approx 0.35$, suggesting that the reference redshift $\zref=0.294$ is incorrect.

\noindent {\it MCXC J0943.1+4659: } SDSS spectra confirm MCXC redshift.

\noindent {\it MCXC J0124.5+0400: } SDSS spectra confirm MCXC redshift.

\noindent {\it MCXC J1447.4+0827: } This cluster has a spectacular star bursting galaxy at its center, with $\zspec=0.375$,
confirming the \redmapper{} redshift.
\\

\noindent {\bf ACCEPT Outliers:}

\noindent {\it MACS J2211.7-0349:} No SDSS spectra, but redshift value in \citet{mantzetal10b} confirms the \redmapper{} redshift.

\noindent {\it Abell 1763:} SDSS spectra confirm \redmapper{} redshift.
\\

\noindent {\bf Mantz Outliers:}

\noindent {\it 400d J0809.6+2811:} This cluster is the same as MCXC J0809.6+2811 (above).  The \redmapper{} redshift
is compromised by bad SDSS photometry from a nearby star.
\\

\noindent {\bf Planck ESZ Outliers:}

\noindent {\it PLCKG56.0-34.9: } This cluster is the same as MCXC J2135.2+0125.  Spectra are unavailable, but 
visual inspection strongly suggests the MCXC redshift is incorrect.


\section{Notes from Centering Analysis}
\label{app:centering_notes}

\noindent {\bf XCS:}

\noindent {\it XCS J2239.4-0547: } This is one of a pair of galaxy clusters at the same redshift, the second being 
XCS J2239.7-0543.  Each XCS system has $T_X=2.8$, but the pair is identified as a single \redmapper{}
cluster.  Since even the X-ray center is ambiguous, we remove these cluster from the centering analysis.

\noindent {\it XMMXCS J1052.4+4419: } There are two clusters near this location at spectroscopic redshift $z=0.44$
and $z=0.50$.  The latter is the XCS system.  \redmapper{} finds both clusters, but the XCS system is heavily masked
by the lower redshift objects, and falls below the detection threshold, so it is difficult to determine whether this system
is a centering failure or not.  We remove the system from our centering study.
\\

\noindent {\bf ACCEPT: }

\noindent {\it Abell 115 (3C 28.0): } The cluster has a N and S component.  \redmapper{} is centered on the S component, but 
ACCEPT centers the cluster on the N component, which dominates the X-ray emission.  
However, both weak lensing \citet{okabeetal10,hurley-walkeretal11}
and dynamical data \citet{barrenaetal07} reveal 2 additional substructures both support the S component
being the dominant one, by a factor of $\approx 4-10$ in mass, which led 
\citet{hurley-walkeretal11} to suggest that the X-ray emission from the N component
is in fact dominated by emission associated with the ratio source 3C 28.
The S component is also clearly dominant in the optical.  Consequently, we assume this clusters
was correctly centered, but caution that our conclusion may be incorrect due to correlated
scatter between optical richness, weak lensing mass, and velocity dispersion.

\noindent {\it Abell 1758: } A1758 has been extensively studied \citep{davidkempner04,okabeumetsu08,hainesetal09,durretetal11b}.
It is typically split into two components --- A1758N and A1758S --- separated by $\approx 2\ \Mpc$.  Both
components are identified as independent clusters by \redmapper, but it is only A1758N which is included in ACCEPT.
A1758N is itself a merging system, with X-ray and optical data suggesting that the N component dominates,
implying \redmapper{} correctly identified the central galaxy of this cluster.  Our central galaxy coincides with the X-ray peak,
but is significantly offset from the X-ray centroid.

\noindent {\it Abell 1914: } This is a complicated merging system with a highly irregular mass distribution \citep{okabeumetsu08,hurley-walkeretal11}.   
Both X-ray and SZ data suggest the brightest cluster galaxy is the center of the dominant component, though the
weak lensing $\kappa$ peak of the NE component appears to be slightly higher (but less extended).  We follow \citet{okabeumetsu08} 
and tentatively associate the SW component as the dominant clump, which makes Abell 1914 a \redmapper{} centering success.

\noindent {\it Abell 370: } There are two comparably bright cD galaxies near the X-ray center of almost equal brightness ($\Delta m =0.05$).   
Contrary to the other times when we were faced with a similar decision, in this case we assigned the dimmer galaxy as the correct center,
based on the curvature of an obvious giant arc.  Our visual choice agrees with the \redmapper{} center.
\\

\noindent {\bf Mantz: }

\noindent {\it MACS J2311.5+0338: } Cluster has distinct NE and SW components.  \redmapper{} centered on the NE, but X-rays indicate
that the SW component is dominant.


\section{Evidence for Spectroscopic Cuts in the WHL Catalog}
\label{app:whl}


\begin{figure}
  \begin{center}
  \epsscale{0.6}
    \plotone{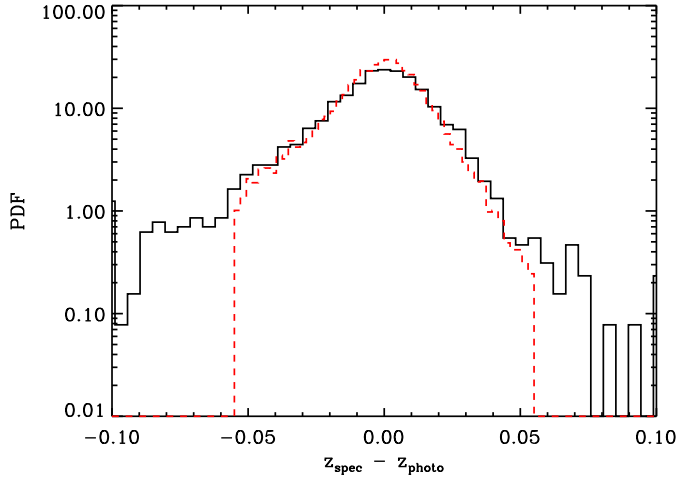}
\caption{Distribution of the redshift offset $z_{spec}-z_{photo}$ for the
\citet{wenetal12} catalog, measured using redshifts exclusive to DR9 (solid 
black histogram) and redshifts from the DR8 release (red histogram).  
All clusters in the redshift range $z_{photo}\in[0.1,0.5]$ are included.
The distribution of the DR8 redshift offsets is clearly artificially truncated
at $|\Delta z| \leq 0.055$, thereby removing the tails of the distribution.
}
\label{fig:whl_compare}
\end{center}
\end{figure}


As noted in section \ref{sec:opt_cats}, we evaluate the redshift performance of the WHL catalog
using clusters with spectroscopic redshifts exclusive to DR9.  Our motivation for excluding all
DR8 \photozs{} from consideration is shown in Figure \ref{fig:whl_compare}, where
we compare the distribution of the redshift offset $z_{spec}-z_{photo}$ as evaluated
using DR8 spectra (red dashed histogram), to that obtained using
redshifts exclusive to DR9 (solid black histogram).
We note that the redshifts exclusive to DR9 were not publicly available at the time 
the \citet{wenetal12} catalog was published.  It is clear from the figure that the distribution
for the DR8 sub-sample is artificially truncated at
$|z_{spec}-z_{photo}| \leq 0.055$.  Given that there is clear evidence this sub-sample
of galaxy clusters was subject to a spectroscopic redshift cut, we evaluate the redshift
performance of the WHL algorithm using only clusters with spectroscopy exclusive to DR9.

\end{document}